\documentclass[accepted=2026-04-15,a4paper,onecolumn,11pt]{quantumarticle}
\pdfoutput=1

\usepackage{comment}
\usepackage{geometry}
\usepackage{cite}
\usepackage{graphicx,amsmath,amssymb,amscd,amsthm}
\usepackage{subcaption}
\usepackage[utf8]{inputenc}
\usepackage[T1]{fontenc}
\usepackage[numbers,square, sort&compress]{natbib}
\usepackage{enumitem}
\usepackage[colorlinks,linkcolor=blue,citecolor=blue,urlcolor=blue]{hyperref}
\usepackage{braket}
\usepackage{xcolor}
\usepackage[normalem]{ulem}
\usepackage{authblk}
\usepackage{siunitx}
\usepackage{bm}

\newcommand{\de}{\delta}

\newcommand{\bla}{\boldsymbol{\lambda}}

\newcommand{\om}{\omega}
\newcommand{\Om}{\Omega}

\newcommand{\iu}{\mathrm{i}}
\newcommand{\dd}{\mathrm{d}}

\newcommand{\id}{1\hspace{-.25em}{\rm I}}

\newcommand{\ie}{i.e.}

\newcommand{\tr}[1]{\text{tr}\left(#1\right)}
\newcommand{\ee}{\mathrm{e}}
\newcommand{\ketbra}[2]{|#1\rangle \! \langle #2|}

\newcommand{\atoms}[1]{{#1}}

\newcommand{\je}[1]{\textcolor{cyan}{#1}}

\title{Holonomic quantum computation: A scalable adiabatic architecture}
\author[1]{Clara Wassner}
\author[1]{Tommaso Guaita}
\author[1]{Jens Eisert}
\author[1]{Jose Carrasco}
\email{jose.carrasco@fu-berlin.de}
\affil[1]{\small Dahlem Center for Complex Quantum Systems, Freie Universit\"{a}t Berlin, 14195 Berlin, Germany}

\begin{document}
\maketitle
\begin{abstract}
    Holonomic quantum computation exploits the geometric evolution of eigen\-spaces of a degenerate Hamiltonian to implement unitary evolution of computational states. In this work we introduce a framework for performing  scalable quantum computation in atom experiments through a universal set of fully holonomic adiabatic gates. Through a detailed differential geometric analysis, we elucidate the geometric nature of these gates and their inherent robustness against classical control errors and other noise sources. The concepts that we introduce here are expected to be widely applicable to the understanding and design of error robustness in generic holonomic protocols. To underscore the practical feasibility of our approach, we contextualize our gate design within recent advancements in Rydberg-based quantum computing and simulation.
\end{abstract}
%%%%%%%%%%%%%%%%%
%%%%%%%%%%%%%%%%%

%%%%%%%%%%%%%%%%%
%%%%%%%%%%%%%%%%%
\section{Introduction}
According to the adiabatic theorem, time evolution under a slowly varying, gapped, time-dependent Hamiltonian preserves spectral subspaces. That is, an eigenstate of the initial Hamiltonian $H(0)$ will evolve to a state in the corresponding eigenspace of the instantaneous Hamiltonian $H(t)$ at later times $t$, provided that this eigenspace is separated by the others by
a uniformly lower bounded
energy gap and that the Hamiltonian varies adiabatically slow in time~\cite{jansen_bounds_2007}. The residual evolution within this eigenspace has a geometric nature~\cite{Zh23}, as initially observed by Berry for non-degenerate Hamiltonians~\cite{Be84}. In this case the only allowed evolution of eigenstates is a phase factor, which exhibits a geometric component known as the Berry phase. This concept naturally extends to Hamiltonians with degenerate spectra: a state in a degenerate eigenspace of the initial Hamiltonian will evolve entirely inside this subspace and undergo a unitary transformation of geometric nature, which can be seen as a generalized Berry phase or a non-Abelian holonomy~\cite{wilczek_appearance_1984}.

In their seminal contribution~\cite{ZR99}, Zanardi and Rasetti have demonstrated the use of non-Abelian holonomies for achieving universal quantum computations. In this framework, the computational space is a degenerate eigenspace
 $V(\bla)$ of a 
family of Hamiltonians $H(\bla)$ parametrized by points $\bla$ within a manifold $M$. Each point $\bla$ represents a configuration of the coupling parameters of the Hamiltonian. Evolving the parameters $\bla(t)$ adiabatically along closed loops
$\bla:[0,1]\to M$ with $\bla(0)=\bla(1)=\bla_0$ induces evolution from the computational space $V(\bla_0)$ into itself. This can be seen as a unitary transformation acting on $V(\bla_0)$. In Ref.~\cite{ZR99}, it is established that, for generic systems, these unitary transformations are sufficient for universal quantum computation. 

In this regard, several examples of implementations of 
single-qubit holonomic gates have been reported~\cite{abdumalikov_jr_experimental_2013,arroyo-camejo_room_2014,zhou_holonomic_2017,CPG20,BTG25} and proposals of two-qubit holonomic gates have 
been formulated for various platforms, including superconducting circuits~\cite{faoro_non-abelian_2003}, trapped ions~\cite{duan_geometric_2001}, quantum dots~\cite{golovach_holonomic_2010, solinas_semiconductor-based_2003}. While each of these platforms has distinct advantages, trapped Rydberg atoms in optical tweezers have recently emerged as a particularly promising system for quantum simulation and quantum computing~\cite{bernien_probing_2017,bluvstein_quantum_2022,cong_hardware-efficient_2022,Levine19,RevModPhys.82.2313}. These architectures enable large-scale programmable quantum simulators~\cite{bernien_probing_2017} and have also demonstrated key ingredients relevant for quantum computing, including the implementation of quantum error-correcting codes and related fault-tolerant primitives~\cite{bluvstein_quantum_2024,GottesmanLDPC,PRXQuantum.2.040101}.
Owing to their intrinsic multilevel structure and strong, controllable interactions, Rydberg-atom platforms are natural candidates for holonomic control. However, despite existing proposals for holonomic single-qudit gates~\cite{GZCKZ22}, a scalable and universal holonomic architecture for this system has so far been lacking.

In this proposal, we complete this picture by introducing a scalable architecture where all gates are realized through adiabatic holonomies. Specifically, we have identified a parametrized Hamiltonian and a universal set of holonomic gates that can be implemented with simple loops in the space of couplings. While the scheme presented is not meant to be a full experimental proposal for a given specific set-up, it should be clear
that the required controlled Hamiltonian is of a form that can be naturally implemented in today's Rydberg atom platforms
\cite{bernien_probing_2017,Levine19,Controlling,PhysRevLett.128.113602,cong_hardware-efficient_2022,bluvstein_quantum_2022,BlochArray,bluvstein_quantum_2024}.

We further study the noise robustness of this protocol to certain types of error, especially in relation to the geometric properties of the system. Owing to their geometric character, holonomic gates are widely regarded as intrinsically robust against small fluctuations in the control parameters~\cite{zhu_geometric_2005,carollo_geometric_2003,de_chiara_berry_2003}. In particular,  to determine the resulting unitary evolution, it is not relevant how exactly the loop is traversed, since only the geometric shape of the loop in parameter space matters. So it is clear that fluctuations in how fast the parameters are varied will not affect the final result.  Similar robustness against these so-called {\em parallel} errors (errors which don't change the shape of the path) has been shown in Ref.~\cite{Barnes21transversal} for cyclic, non-degenerate Hamiltonian evolutions. In the geometric language, this feature is a consequence of the invariance of certain integrals under diffeomorphisms, or, simply put, a change of variables in the integration. For what instead concerns {\em perpendicular} errors, i.e., errors that slightly deform the shape of the loop, the intuition is less clear. In this
work, we will argue that resistance to this type of errors is intrinsically related to another geometric notion, namely the \emph{curvature} of the vector bundle represented by the manifold $M$ and the vector spaces $V(\bla)$. We observe that this curvature can be made arbitrarily small in our protocols, leading to an enhanced robustness to coherent errors of both types discussed above. 

We believe that this latter analysis is of value for clarifying the true nature of holonomic error resistance, even for cases beyond the current specific implementation. This geometric picture also allows us to draw a parallel between the noise resilience of holonomic gates and the one of fault tolerant computational models, which can also be seen as relying on the existence of a vector bundle with zero curvature~\cite{gottesman2017fibre}. 

\atoms{In a way, our proposal can be seen as being comparable to the results presented in Refs.~\cite{Levine19} or~\cite{GZCKZ22}, where Rydberg 
interactions and the ability to manipulate atoms using tweezers are exploited to implement entangling gates between individual atoms.} 
However, it is crucial to note that our protocols differ significantly from those described in the aforementioned references. Firstly, all our gates are fully holonomic, benefiting from an ensuing robustness to certain errors, as we will display in this work. Second,
a larger class of single-qubit and two-qubit gates can be implemented naturally within our framework. Third, our protocol is not limited to qubit systems but is generalised to the higher dimensional qudit case in a trivial way, making it relevant for instance for the simulation of lattice theories with high dimensional local spaces~\cite{LatticeGauge,GZCKZ22}.

The 
remainder
of this 
work is structured as follows. In Sec.~\ref{sec:summary} we give a high level description of the constructions we introduce and of the results that we derive. More concretely, the general set up of the architecture is presented and the universal set of gates that can be achieved holonomically is introduced. In Sec.~\ref{sec:methods} the detailed derivation of 
these results is explained, making use of the language of differential geometry. In particular, in Sec.~\ref{sec:methods-errors} the robustness to errors of the holonomic gates is analyzed in detail. Finally, in Sec.~\ref{sec:discussion} the implications
for actual physical architectures and 
an outlook for this protocol are discussed.

\section{Summary of main results}\label{sec:summary}
In this section, we concisely present the main findings and structure of our work.

\paragraph{Introduction of our Hamiltonian.}
%Guided by but not restricted to the above physical picture, 
In this work we consider a model in which each logical qudit is encoded in a $(d+2)$-level system with basis states $\ket{0}, \dots, \ket{d}, \ket{\mathrm{f}}$ and assume that the effective single-particle Hamiltonian governing the evolution % in the interaction picture 
is of the form %(see Appendix~\ref{sec:app_Hphysical})
\begin{equation}\label{eq:1qubit_hamiltonian}
    H_0(\bla)=\sum_{a=0}^d \Omega_a \ket{a} \bra{\mathrm{f}} +\mathrm{h.c.}\,,
\end{equation}
where $\{\Omega_a\}_{a=0}^d$ are tunable, complex parameters. In the remainder of our work we will assume that $\Omega_d$ is fixed at some non-zero value and that the other parameters shift adiabatically by suitable classical control. These controllable parameters are the real and imaginary parts of $\Om_0,\ldots,\Om_{d-1}$ which we shall collectively denote as $\bla$. For two identical systems of this type, we introduce the interacting Hamiltonian
\begin{equation}
    H(\bla)=1\otimes H_0(\bla)+H_0(\bla)\otimes1+ W \ket{d,d}\bra{d,d}\,, \label{eq:2-atom-hamiltonian}
\end{equation}
where $H_0(\bla)$ is the single-system Hamiltonian given by Eq.~\eqref{eq:1qubit_hamiltonian} and $W$ is an interaction strength that depends on the system settings. In particular, we assume that there exists a setting where $W=0$ and a setting where $W$ is a finite positive value.

Holonomic quantum computations will be conducted in the zero-energy subspaces of the aforementioned Hamiltonians. We identify zero-energy 
eigenstates at each point $\bla$ and provide explicit analytic expressions for them. Specifically, when $W=0$ in Eq.~\eqref{eq:2-atom-hamiltonian}, these subspaces exhibit a tensor product structure of two $d$-dimensional subspaces, each linked to the zero-energy eigenstates of the Hamiltonian $H_0(\bla)$ from Eq.~\eqref{eq:1qubit_hamiltonian}. However, when $W\neq0$, a $d^2$-dimensional subspace emerges that is not a tensor product, enabling the implementation of two-qudit entangling gates.

These Hamiltonians are inspired by structures that naturally appear in several quantum platforms where multilevel systems are coherently driven and where interactions between selected excited levels can be effectively switched on and off.
Neutral atoms coupled through Rydberg excitations constitute a prominent example of such a setting: long-lived electronic states provide a qudit manifold, optical fields induce effective Rabi couplings $\Omega_a$ (often via two-photon transitions), and bringing atoms together in optical tweezers gives rise to a strong, tunable interaction between simultaneous Rydberg excitations~\cite{Levine19,PhysRevLett.128.113602,Controlling,BlochArray,bluvstein_quantum_2024,Weimer}. In this Rydberg scenario, the doubly excited state $\ket{d,d}$ undergoes a distance-dependent energy shift that can be approximated by a soft-core blockade potential with a van-der-Waals tail, leading—under strong blockade—to an effective energy penalty $W$~\cite{Weimer}. However, we stress that our holonomic construction does not rely on the detailed microscopic form of this interaction: the only essential feature is the existence of two operational regimes, one in which the effective parameter $W$ is negligible ($W\simeq0$) and another in which it is finite and positive ($W>0$). Neutral-atom tweezers provide a compelling physical realization of such a tunable interaction~\cite{bluvstein_quantum_2024}, where the interacting regime corresponds to atoms brought close together and the non-interacting one corresponds to atoms positioned far apart compared to the blockade radius.
Alternative physical platforms with equivalent controllability would, however, be equally suitable.

For completeness, Appendix~\ref{sec:app_Hphysical} provides a compact derivation showing how the single-atom interaction-picture Hamiltonian~\eqref{eq:1qubit_hamiltonian} follows from standard approximations in driven atomic systems (dipole approximation and rotating-wave approximation), though the subsequent analysis does not depend on this particular realization.

\paragraph{Differential geometry.}
To study holonomic quantum computation with Hamiltonians like the one introduced 
above, it is useful to introduce some concepts of differential geometry. This perspective allows us to describe the system with a concise language.  It also allows us to develop a more general intuition of what is happening, that goes beyond the specific example Hamiltonian we are considering here. In Sec.~\ref{sec:methods-diff} we will therefore introduce the basic framework of holonomic quantum computing in differential geometry terms.

We start by observing that the full set of Hamiltonians which we consider can be seen as a manifold $M$. Each point of the manifold is a Hamiltonian $H(\bla)$, parametrized by a certain set of couplings and configurations that we collect in the vector $\bla$. In our proposal, the computational space is defined as the null subspace of $H(\bla)$ for each choice of $\bla$, which we indicate as $V(\bla)$. This structure that emerges is a vector bundle on $M$: that is, essentially, a manifold where at each point $\bla$ a vector space $V(\bla)$ is attached in a continuous fashion. This vector bundle is endowed with a natural notion of curvature, represented by a connection 1-form $A$, whose matrix elements can be expressed in terms of the eigenvectors of the Hamiltonians in the manifold. The notion of curvature in turn introduces a natural way to define the parallel transport of vectors between the different vector spaces $V(\bla)$ that compose the vector bundle. 

Consider now the adiabatic evolution of the Hamiltonian couplings along a closed path $\bla(t)$ in the manifold, with $t\in[0,T]$ and $\bla(0)=\bla(T)=\bla_0$. The adiabatic condition implies that states confined in the null subspace of the initial Hamiltonian will remain in the null subspaces of the subsequent time-dependent Hamiltonians. That is, if $\ket{\psi(0)}\in V({\bla}_0)$, then $\ket{\psi(t)}\in V({\bla}(t))$ for all times $t$ as long as ${\bla}(t)$ changes adiabatically. At time $t=T$ the state must return to the initial null subspace, possibly rotated by some unitary transformation. That is $\ket{\psi(T)}=U\ket{\psi(0)} \in V({\bla_0})$.
We will see that, from the geometric perspective, this adiabatic evolution of the state coincides exactly with the parallel transport along the vector bundle, as defined in terms of the connection $A$. This means that the transformation of $\ket{\psi(t)}$ is entirely captured by the geometrical curvature of the bundle. Indeed, both the time-evolved state vector $\ket{\psi(t)}$ and the final unitary transformation $U$ can be expressed in terms of path-ordered integrals of the connection which we derive Sec.~\ref{sec:methods-diff}.

\paragraph{Analytical solution of the parallel transport equations: a universal set of holonomic gates.} 

\begin{figure}
    \centering
        \begin{subfigure}[b]{0.32\textwidth}
        \centering    
        \includegraphics[width=\textwidth]{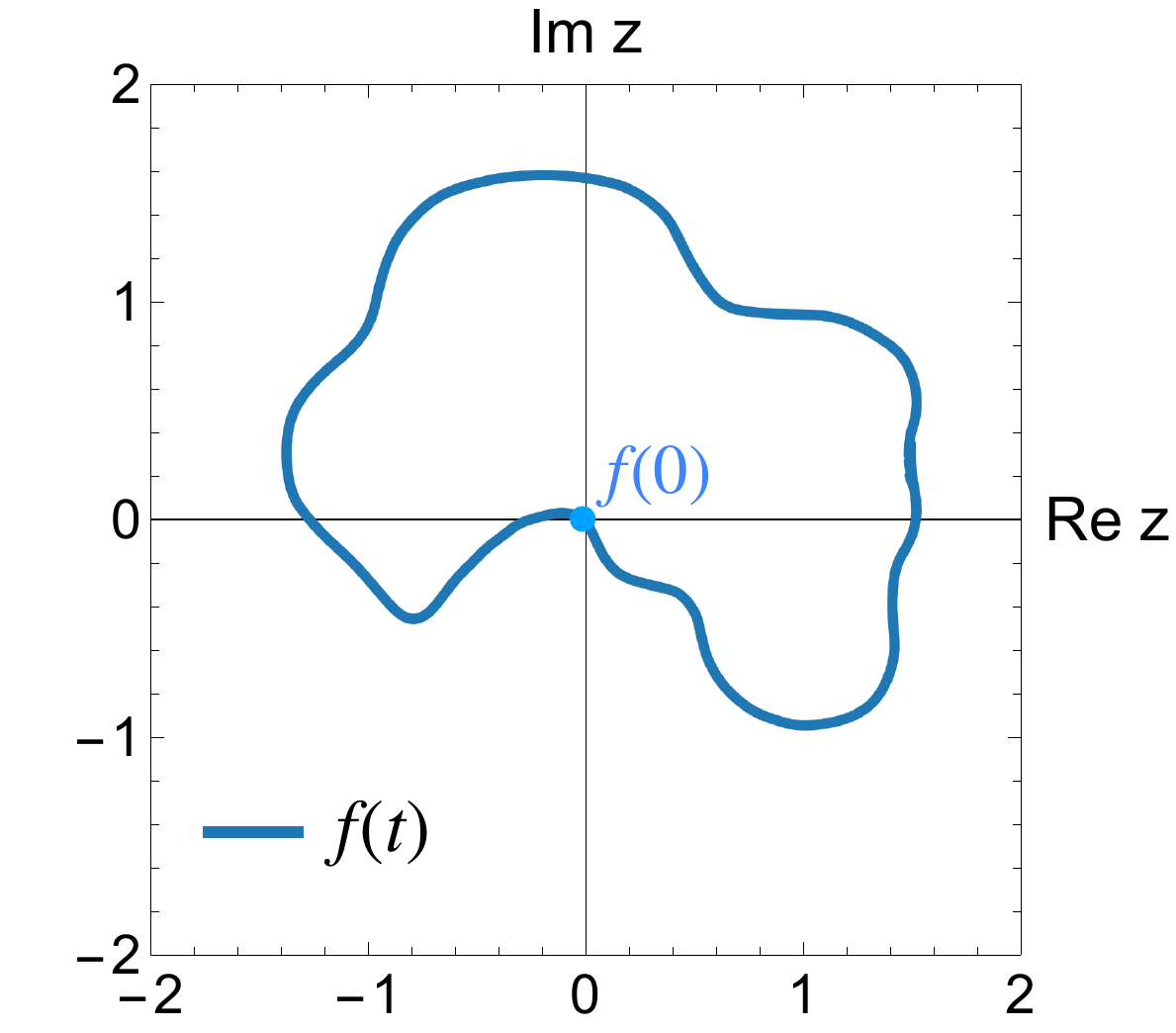}
        \caption{}
        \label{fig:general_loop_integrands-a}
    \end{subfigure}
    \hfill
    \begin{subfigure}[b]{0.32\textwidth}
        \centering
     \includegraphics[width=\textwidth]{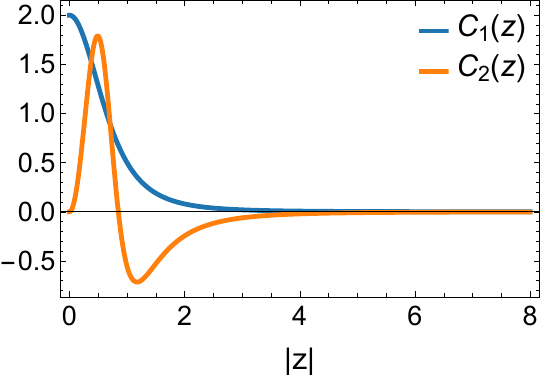}
        \caption{}
        \label{fig:general_loop_integrands-b}
    \end{subfigure}
    \hfill
        \begin{subfigure}[b]{0.32\textwidth}
        \centering
        \includegraphics[width=\textwidth]{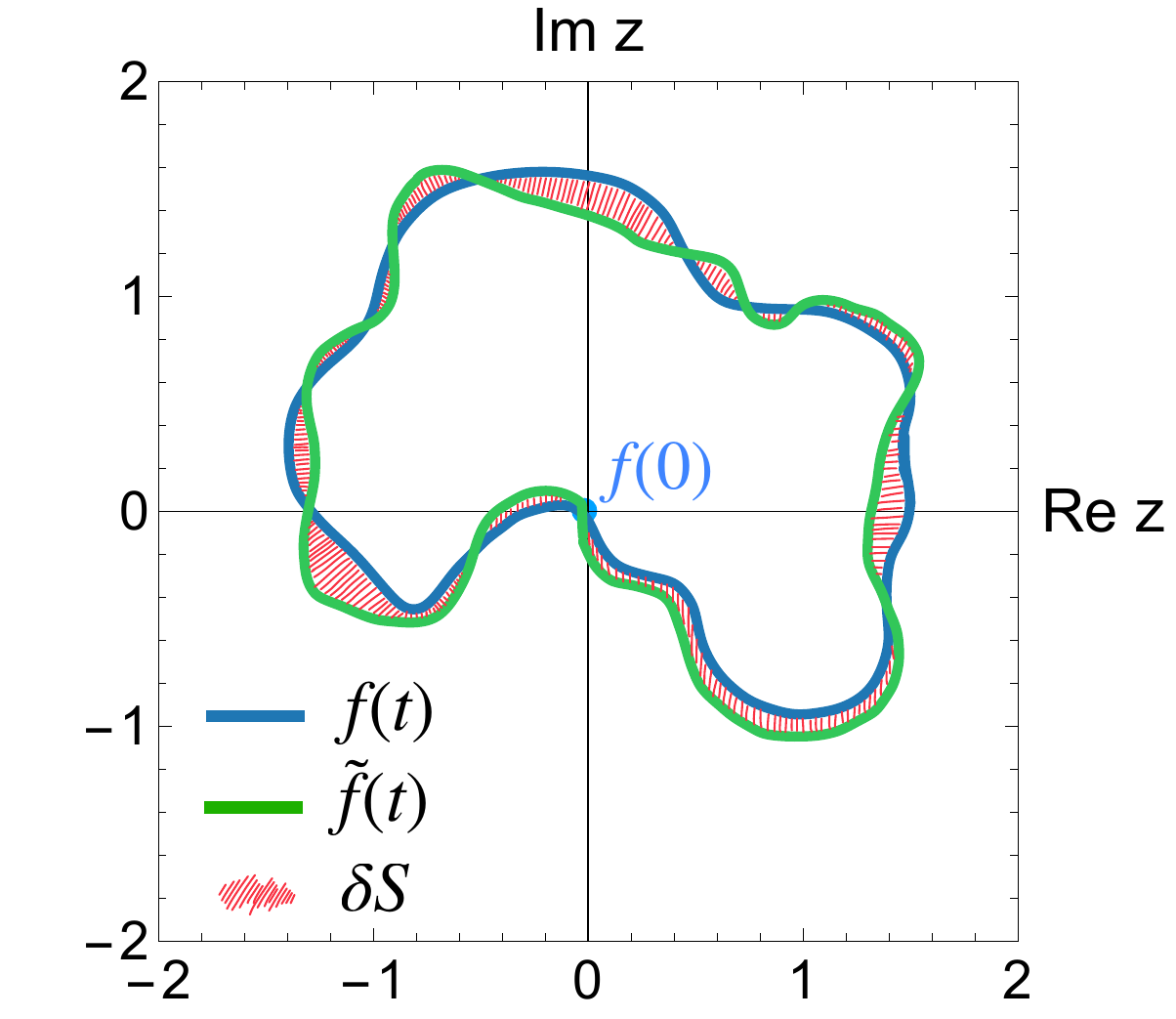}
        \caption{}
        \label{fig:general_loop_integrands-c}
    \end{subfigure}
    \caption{(a) For an arbitrary loop $f(t)$ in the complex plane, as for example depicted here, we set the transition amplitudes (our Hamiltonian parameters) to be $\mathbf{\Omega}(t)=|\Omega_d|f(t) \bm{\omega}$, with $\bm{\omega} \in \mathbb C^d$ an arbitrary, constant unit-vector. This allows us to analytically determine the unitary implemented by an adiabatic evolution along this loop. (b) The analytic expressions for the implemented unitaries depend on two phases where each is the result of a surface integral over the area enclosed by $f$.
    %$f(t)$. 
    The two spherically symmetric integrands of the two surface integrals which determine the phases $\alpha_1$ and $\alpha_2$ are shown here.
    (c) Coherent errors can be interpreted as a deformed loop $\tilde f$.
    %$\tilde f(t)$. 
    The impact of the error is directly related to the surface integral over the excess area $\delta S$.}
\end{figure}

In general, the path-ordered integral which determines the implemented unitary transformation $U$ cannot be computed analytically. However, for the Hamiltonian introduced above, we identify a set of loops where we can assign to every loop the corresponding unitary transformation 
in an analytical fashion. More concretely, we choose the complex transition amplitudes to be of the form $\Omega_a(t)=|\Omega_d|f(t)\,  \omega_a$ for all $0\le a\le d-1$ with $\bm{\omega} \in \mathbb C^d$ an arbitrary constant unit-vector and $f(t) \in \mathbb C$ any closed loop in the complex plane with $f(0)=f(T)=0$ (see Fig.~\ref{fig:general_loop_integrands-a}).
Letting every transition amplitude $\Omega_a(t)$ evolve along the same loop (up to a constant factor $\omega_a$) permits us to solve the parallel transport condition analytically to obtain the implemented unitary. Our first main result, that we show in Sec.~\ref{sec:methods-phase-sols}, states that when the previous path is followed adiabatically the %following 
single- and two-qudit quantum gates
\begin{equation}\label{eqs:our_gate_1and2qubits}
\begin{aligned}
U_{(1)}&=\exp\big[\iu\alpha_1\,|\omega \rangle \langle \omega|\big]\,,\\
U_{(2)}&=\exp\big[\iu\alpha_1(1 \otimes |\omega \rangle \langle \omega|+ |\omega \rangle \langle \omega| \otimes 1) \big]\cdot\exp\big[\iu\alpha_2(\ket{\omega}\otimes \ket{\omega})(\bra{\omega}\otimes \bra{\omega})\big]\,,
\end{aligned}
\end{equation}
are implemented,
with $\ket{\omega}=\sum_a \omega_a \ket{a}$.
The real numbers $\alpha_1,\alpha_2$ result from surface integrals over the area $S$ enclosed by the loop $f(t)$:
\begin{equation}\label{eqs:alphas}
\begin{aligned}
\alpha_1&=\int_S \dd^2z  \;\mathcal C_1(z),\\
    \alpha_2&=\int_S \dd^2z  \; \mathcal C_2(z) .
\end{aligned}
\end{equation}
The spherically symmetric functions $\mathcal C_1,\mathcal C_2: \mathbb C \rightarrow \mathbb R$ are defined and discussed in detail in Sec.~\ref{sec:methods-phase-sols} (see Eqs.~\eqref{eq:C1} and~\eqref{eq:C2} below for explicit formulas)
and plotted in Fig.~\ref{fig:general_loop_integrands-b}.
For $\bm{\omega} = (0,1)^{\text{T}}$ the loop $f(t)$ can be chosen such that the two-qubit gate $U_{(2)}$ is equivalent to the entangling controlled-$Z$ gate on two qubits up to single-qubit rotations. Furthermore, we can implement any single-qubit gate by suitably choosing $f(t)$ and $\bm{\omega}$. This means that we have identified a universal set of holonomic gates each implementable with $O(1)$ loops in the parameter space. 

\paragraph{Error analysis.}
In Sec.~\ref{sec:methods-errors}, we will argue that a holonomic implementation like ours possesses an intrinsic resistance to coherent errors in the Hamiltonian parameters. We show this by rigorously analyzing the impact of errors on the specific one- and two-qudit gates that we propose. Beside this, we also highlight how this resistance arises from general geometric considerations, which we expect to hold also for other holonomic protocols. In this sense, our work clarifies the main aspects of the widely held belief that holonomic gates are more robust to errors and makes an intriguing connection to the geometric structures that are expected lie at the base of fault tolerant models of computation~\cite{gottesman2017fibre}. %In a way, our holonomic protocol implements, in an approximate fashion, some of the key features that enable fault tolerant computation.

More in detail, the coherent errors that we consider are due to the fact that the Hamiltonian couplings $\bla(t)$ implemented along the adiabatic loop are not exactly the correct ones, but rather some distorted version. In other words, we generate the adiabatic evolution associated not to the desired loop $\bla(t)$ but a deformed loop $\tilde \bla(t) $. We have seen that our protocol produces gates that are essentially phase gates, with phases determined by the expressions $\alpha_1$, $\alpha_2$ above. An error in the loop will generate an error in the value of these phases, which will in turn lead to faulty gates. 

Notice, however, that the way in which $\alpha_1$, $\alpha_2$ depend on the loop $\bla(t)$ is through a surface integral on the region enclosed by the loop, as in Eq.~\eqref{eqs:alphas}. Errors which, for instance, only change the velocity with which the loop is followed, will not lead to any change to this region and thus to the gate. Errors which, on the other hand, change the exact shape of the loop $f(t)$ in the complex plane will lead to a variation in $\alpha_1$, $\alpha_2$ bounded by
\begin{equation} \label{eq:delta-alpha}
    \delta \alpha \leq \int_{\delta S} d^2 z \, \big|\mathcal{C}(z)\big| \;\leq\; \left|\delta S\right| \, \max_{z\in \delta S} \big|\mathcal{C}(z)\big|\,,
\end{equation}
where $\delta S$ is the region of the complex plane in between the correct and the distorted loop (see Fig.~\ref{fig:general_loop_integrands-c} for an example), and $\left|\delta S\right|$ is the area of this region. This area expresses essentially the magnitude of the error. 
The further the distorted loop lies from the correct one, the greater this area 
will be. 

By examining Eq.~\eqref{eq:delta-alpha} we see that, as expected, the deviation from the ideal gate depends on this error magnitude $\left|\delta S\right|$. However, the proportionality coefficient of this dependence is related to the value of the function $\mathcal{C}$ along the loop. We have seen (Fig.~\ref{fig:general_loop_integrands-b}) that this function decays rapidly to zero away from the origin of the complex plane (both in the case of the one- and of the two-qudit gates). So if we choose a loop that mostly runs in regions sufficiently far away from the origin, the impact of coherent errors on the implemented gates will be correspondingly suppressed. This can always be achieved, except for the initial and final parts of the loop, which must close at the origin: these parts will thus be the most susceptible to possible errors.

From a higher level point of view, a small deformation of a path on a vector bundle leads to a change in the parallel transport that is proportional to the curvature of the bundle where this distortion happens. The vector bundle must have some regions of non-trivial curvature, such that there exist some closed loops that lead to non-trivial parallel transport (and thus non-trivial gates). However, in cases like ours it is possible to find loops that mostly cross regions of low curvature: if an error occurs in these regions, its impact on the final outcome is negligible.  

The robustness of the scheme is primarily against control errors in the path. That said,
for
%For 
completeness, in Sec.~\ref{sec:methods-errors-decoherence} we also study the impact of other forms of errors, such as decoherence or approximations in the adiabatic condition, which cannot be described purely as deformation of the parameter loop. In Sec.~\ref{sec:incoherent},
we will further sketch how the robustness to coherent errors, as given by the upper bound Eq.~\eqref{eq:delta-alpha}, may also apply to some incoherent error models, when these can be represented as classical stochastic mixing of coherent paths.

\section{Derivations of main results}\label{sec:methods}
\subsection{Differential geometry}\label{sec:methods-diff}
As already mentioned previously, the transformations that arise during an adiabatic loop are intrinsically geometric in nature. More specifically, the whole process can be understood using the language of differential geometry in the following way. The Hamiltonian~\eqref{eq:2-atom-hamiltonian} can be interpreted as a parametrized Hamiltonian $H(\boldsymbol{\lambda})$, where the vector of real parameters $\boldsymbol{\lambda}$ represents a specific configuration of the variable couplings in the Hamiltonian. We will indicate the individual components of $\bla$ as $\lambda^\mu$. In our case, $\boldsymbol{\lambda}$ corresponds to $2d$ real parameters defining the real and imaginary parts of $\Omega_a$ for $0\le a\le d-1$. The collection of the Hamiltonians $H(\boldsymbol{\lambda})$ for all $\bla$ defines a parametrized manifold $M$.
On this base manifold it is possible to define a vector bundle. Indeed, for each choice of parameters $\boldsymbol{\lambda}$, we can identify a vector space $V(\boldsymbol{\lambda})$ given by the span of the null eigenstates of $H(\boldsymbol{\lambda})$. The collection of these $(d^2+1)$-dimensional vector spaces defines a vector bundle on the manifold $M$.

Consider now the situation where the Hamiltonian is evolved in time along a certain path $\bla(t)$ in parameter space. If a state is initially in the null eigenspace of the initial Hamiltonian, changing the parameters adiabatically means that the state will remain in the same eigenspace of the time-evolved Hamiltonian, that is, it will remain within the vector bundle. More precisely, if $\ket{\psi(0)}\in V(\boldsymbol{\lambda}(0))$ then $\ket{\psi(t)}\in V(\boldsymbol{\lambda}(t))$ for all $t$, if $\boldsymbol{\lambda}(t)$ changes adiabatically slowly (see Theorem 3 of Ref.~\cite{jansen_bounds_2007} for a precise statement of the adiabatic condition). If a state vector $\ket{\psi(t)}$ does follow such an adiabatic evolution, it holds that
\begin{equation}
    \frac{d}{dt}\ket{\psi(t)} = -iH(\bla(t))\ket{\psi(t)} = 0\,.
    \label{eq:adiabatic-time-evo}
\end{equation}
Indeed, $H(\bla(t))\ket{\psi(t)}$ will vanish if $\ket{\psi(t)}$ is in the null eigenspace of the Hamiltonian at all times.\footnote{Note that the more general case, where the encoding eigenspaces have a non-zero eigenvalue, can be reduced to this case by adding a constant shift to the Hamiltonian. This will induce a global dynamical phase of non-geometric nature in the evolution of all states, which can be straightforwardly kept track of. For simplicity we will focus here on zero energy eigenspaces.}

On a vector bundle it is possible to define a notion of covariant derivative, that is a map $\nabla$ on the vector spaces satisfying certain specific conditions that allow it to be interpreted as a derivative (see, for example, Ref.~\cite{nakahara_geometry_2003} for more details). In our case, a natural choice is 
\begin{equation}
    \nabla_\mu \ket{ \chi(\boldsymbol{\lambda})} :=\mathbb{P}_{V(\boldsymbol{\lambda})} \: \frac{\partial}{\partial\lambda^\mu} \ket{\chi (\boldsymbol{\lambda})} \,%.
    ,
    \label{eq:covariant_deriv_def}
\end{equation}
where $\ket{\chi(\bla)}$ is a vector in $V(\bla)$ for all $\bla$ and $\nabla_\mu$ indicates the covariant derivative with respect to the $\mu$-th parameter coordinate $\lambda^\mu$. Here $\mathbb{P}_{V(\bla)}$ is the orthogonal projector on the subspace $V(\bla)$. The covariant derivative induces a local notion of parallel transport of vectors among the vector spaces of the vector bundle. More specifically, a vector $\ket{\chi(\bla(t))}$, defined in the vector spaces corresponding to a path $\bla(t)$, is said to be parallel transported if
\begin{equation}
    \dot\lambda^\mu  \nabla_\mu \ket{ \chi(\bla(t))} =  0 \,,
\end{equation}
for $\dot\lambda^\mu=\frac{d}{dt}\lambda^\mu(t)$. Notice that here and in what follows we have used Einstein's convention that sums over repeated indices are assumed. It follows that the adiabatic time evolution equation~\eqref{eq:adiabatic-time-evo} can thus be interpreted as a parallel transport condition according to the natural covariant derivative defined above. Indeed,
\begin{equation}
    0= \mathbb{P}_{V(\bla(t))} \frac{d}{dt}\ket{\psi(t)} = \frac{d\lambda^\mu}{dt} \, \mathbb{P}_{V(\bla(t))} \, \frac{\partial}{\partial\lambda^\mu}\ket{\psi(\bla(t))} =  \dot\lambda^\mu \nabla_\mu \ket{\psi(\bla(t))}\,.
    \label{eq:parallel-transport-derivation}
\end{equation}

\begin{sloppypar}Fixing local vector bases $\{\ket{v_a(\boldsymbol{\lambda})} \}$ for each space $V(\boldsymbol{\lambda})$, one can expand the time evolving vector according to $\ket{\psi(t)}:= \psi^a(t)\, \ket{v_a(\boldsymbol{\lambda}(t))}$. 
Notice that we have not required the basis vectors $\ket{v_a(\boldsymbol{\lambda})}$ to be orthonormal, so they will in general have a non-trivial Gram matrix ${g_{ab}:=\braket{v_a(\boldsymbol{\lambda})|v_b(\boldsymbol{\lambda})}}$. We may, therefore, 
write the subspace projector as ${\mathbb{P}_{V(\boldsymbol{\lambda})}=\ket{v_a(\boldsymbol{\lambda})}g^{ab}\bra{v_b(\boldsymbol{\lambda})}}$, where $g^{ab}$ is defined as the inverse of $g_{ab}$, \ie, 
$g^{ab}g_{bc}=\delta^a_c$. Using this we can express Eq.~\eqref{eq:covariant_deriv_def} in the basis coordinates as\end{sloppypar}
\begin{equation}
\begin{aligned}
     \nabla_\mu \chi^a(\boldsymbol{\lambda}) &= \partial_\mu \chi^a(\boldsymbol{\lambda}) + g^{ab} \braket{v_b(\boldsymbol{\lambda})|\partial_\mu v_c(\boldsymbol{\lambda})} \chi^c(\boldsymbol{\lambda})\\
     &= \partial_\mu \,  \chi^a(\boldsymbol{\lambda}) + A^a{}_{c\mu}(\boldsymbol{\lambda}) \: \chi^c(\boldsymbol{\lambda})\,,
\end{aligned}
\end{equation}
where we have introduced the connection 1-form tensor $A^a{}_{b\mu}(\boldsymbol{\lambda}) := g^{ac} \braket{v_c(\boldsymbol{\lambda})|\partial_\mu v_b(\boldsymbol{\lambda})}$. This should be interpreted as a non-Abelian Berry connection. We will sometimes also use the connection with lowered indices $A_{ab\mu}=\braket{v_a(\boldsymbol{\lambda})|\partial_\mu v_b(\boldsymbol{\lambda})} = g_{ac} A^c{}_{b\mu}$. With this notation, the parallel transport condition~\eqref{eq:parallel-transport-derivation} for $\ket{\psi(t)}$ takes the form 
\begin{equation}
    \frac{d}{dt} \psi^a(t) + \dot{\lambda}^\mu A^a{}_{b\mu}(\boldsymbol{\lambda}(t)) \: \psi^b(t) =0 \,,
    \label{eq:parallel-transport}
\end{equation}
whose formal solution is given by the time-ordered exponential
\begin{equation}
    \psi(t) = \underbrace{\mathcal{T} \exp \left[ -\int_0^t  \dot{\lambda}^\mu A_{\mu}(\boldsymbol{\lambda}(t')) \: dt' \right]}_{U(t)} \psi(0)\,, \label{eq:time-ordered-int}
\end{equation}
where $A_\mu$ is just the matrix obtained by considering the indices $a$ and $b$ of $A^a{}_{b\mu}$ for a fixed $\mu$.

In summary, the vector bundle that we are considering is endowed with a natural notion of parallel transport, encoded in the concept of covariant derivative. Once a specific choice of basis vectors has been made, the connection tensor $A^a{}_{b\mu}$ contains all the necessary information to reconstruct the covariant derivative (and thus parallel transport) and its entries can be expressed concisely in terms of the vector basis elements. If the parameters $\bla(t)$ are evolved adiabatically in time, then a state vector $\ket{\psi(t)}$ will be parallel transported along the vector bundle, satisfying Eq.~\eqref{eq:parallel-transport}. The map $U(t)$ defined in Eq.~\eqref{eq:time-ordered-int}, which formally solves the parallel transport equation, shows how a vector initially in $V(\bla(0))$ is parallel transported to a vector in $V(\bla(t))$.
Consider now the case in which, after a time $T$, the parameters $\boldsymbol{\lambda}(t)$ complete a closed loop, that is $\boldsymbol{\lambda}(0)=\boldsymbol{\lambda}(T)$. Then the map $U(T)$ %, defined by the path-ordered integral above,
is a unitary transformation from $V(\boldsymbol{\lambda}(0))$ onto $V(\boldsymbol{\lambda}(T))$, that is, onto itself. It follows that $U(T)$ performs a unitary gate on the information encoded in the subspace $V(\boldsymbol{\lambda}(0))\equiv V(\boldsymbol{\lambda}(T))$.

If a vector $\ket{\psi}$ is transported along two different paths $\bla(t)$ and $\bla'(t)$, one of which is a slight deformation of the other, the outcome will be in general two slightly different vectors. The magnitude of this difference is related to a property of the connection known as \emph{curvature}. We will argue later that some of our results can be interpreted as the fact that our manifold in some regions has approximately vanishing curvature, i.e., different paths lead to the same parallel transported vector.  

\subsection{Analytic parametrization of phase gates}\label{sec:methods-phase-sols}

In this section, we analyze the connection $A$ and the unitary $U(T)$ in two cases: (i) when the adiabatic loop is carried out on an isolated atom governed by the Hamiltonian~\eqref{eq:1qubit_hamiltonian}, and (ii) when it is carried out on two atoms that are nearby and interact according to Hamiltonian~\eqref{eq:2-atom-hamiltonian} with $W>0$. We derive in particular the explicit expressions~\eqref{eqs:our_gate_1and2qubits}
and~\eqref{eqs:alphas}, which are the solutions of Eq.~\eqref{eq:time-ordered-int} for loops of the form $\Omega_a(t)=|\Omega_d|f(t) \, \omega_a$ for $0\le a\le d-1$.

\subsubsection{Single-qudit unitaries}\label{sec:methods-1qudit}
%Here 
In this section, we derive the unitary $U(T)$ when the adiabatic loop is carried out on an isolated atom governed by Hamiltonian~\eqref{eq:1qubit_hamiltonian}. In this case, the null subspaces of the Hamiltonian~\eqref{eq:1qubit_hamiltonian} are $d$-dimensional 
and can be characterized as follows. First, it is easy to show 
that the vectors
\begin{equation}
 \ket{\pm}=\Omega \ket{\mathrm{f}} \pm \sum_{a=0}^d \Omega_a \ket{a}
\end{equation}
are (unnormalised) eigenvectors
of $H_0(\bla)$ with eigenvalue $\pm\Om$, where $\Om^2=\sum_{a=0}^d|\Om_a|^2$. Second, the vectors
\begin{equation}\label{eq:null-ev-1qudit}
    \ket{e_a(\bla)}=\overline{\Omega}_a\ket d-\overline{\Omega}_d\ket a\,,
\end{equation}
for all $0\le a\le d-1$, are linearly independent and belong to the null subspace of $H_0(\bla)$. Remember that we collect in the vector $\boldsymbol{\lambda}$ the real parameters of the model, i.e., $\boldsymbol{\lambda}=(\mathrm{Re}\, \Omega_0,\dots, \mathrm{Re}\, \Omega_{d-1},$ $\mathrm{Im} \,\Omega_0,\dots, \mathrm{Im} \,\Omega_{d-1})$. The parameter $\Omega_d$ is kept at a fixed finite value.

The Hamiltonian $H_0(\bla)$ thus has a $d$-dimensional null state manifold, separated from the other energy eigenstates by a gap $\Omega\geq\Omega_d$. This subspace will be the local computational subspace where the logical information is encoded. Notice that at the point $\bla_0$ where $\Omega_a=0$ for all $0\le a\le d-1$, which we will consider as the base point for all our transformations, this computational subspace is given simply by the span of the levels $\ket{0}, \dots, \ket{d-1}$. 

Given that the null subspaces $V(\bla)$ are spanned by the basis vectors $\{\ket{e_a(\boldsymbol{\lambda})}\}$ defined in Eq.~\eqref{eq:null-ev-1qudit}, we can straightforwardly compute the quantities introduced previously in Sec.~\ref{sec:methods-diff}. More specifically, we have
\begin{equation}\label{eq:g-1qudit}
g_{ab}(\bla):=\braket{e_a(\boldsymbol{\lambda})|e_b(\boldsymbol{\lambda})}= |\Omega_d|^2\left(\delta_{a\je{,}b} + \frac{\Omega_a \overline{\Omega}_b}{|\Omega_d|^2} \right)
\end{equation}
for all $0\le a,b\le d-1$. Therefore, writing $\Omega^2=\sum_{a=0}^d |\Omega_a|^2$ as before,
\begin{equation}\label{eq:gi-1qudit}
g^{ab}(\bla)=|\Omega_d|^{-2}\left(\delta_{a\je{,}b} - \frac{\Omega_a \overline{\Omega}_b}{\Omega^2} \right)
\end{equation}
for all $0\le a,b\le d-1$, 
are the matrix elements of the inverse $g^{-1}$ of $g$. It also follows that
\begin{equation}\label{eq:A-1qudit}
A_{abc}(\bla):=\braket{e_a(\boldsymbol{\lambda})|\partial_c e_b(\boldsymbol{\lambda})}=\begin{cases}
    \Omega_a \delta_{c,b},&0\le c\le d-1,\\
    -\iu\Omega_a \delta_{c-d,b},&d\le c\le 2d-1
\end{cases}
\end{equation}
for all $0\le a,b\le d-1$. The first and second lines correspond to taking derivatives with respect to the real and imaginary parts of $\Omega_c$  respectively. Combining these results\je{,} 
we arrive at
\begin{equation}
    \dot{\lambda}^\mu A^a{}_{b\mu}(\bla)= \frac{\Omega_a}{\Omega^2}(\mathrm{Re}\, \dot{\Omega}_b -i\, \mathrm{Im} \,\dot{\Omega}_b) = \frac{\Omega_a \dot{\overline{\Omega}}_b}{\Omega^2}
\end{equation}
for all $0\le a,b\le d-1$. %, which in vector notation corresponds to $\dot{\lambda}^\mu A_{\mu}=\Omega^{-2}\,\Omega\dot{\Omega}^\dag$. 
Assume now that we adiabatically move the parameters along the adiabatic loop defined by $\Omega_a(t)=|\Omega_d|f(t)\omega_a$ for $0\le a\le d-1$, where $f(t)$ is a complex function such that $f(0)=f(T)=0$ and $\sum_a |\omega_a|^2=1$. We then have that 
\begin{equation}
    \dot{\lambda}^\mu A^a{}_{b\mu}(t) =  |\Omega_d|^2 \Omega^{-2}(t) \, f(t){\dot{\bar f}}(t)\, \omega_a \overline{\omega}_b\,,
\end{equation}
with $\Om^2(t)=|\Om_d|^2(1+f(t)\bar f(t))$, which in particular implies that the matrices $\dot{\lambda}^\mu A_{\mu}(t)$ commute with each other at different times. This simplifies the evaluation of the time-ordered integral~\eqref{eq:time-ordered-int} which reduces to
\begin{equation}
    U(T)=\mathcal{T} \exp \left[ -\int_0^T  \dot{\lambda}^\mu A_{\mu}(\boldsymbol{\lambda}(t))\,\dd t \right]
    =\exp \left[ -\int_0^T  \frac{f(t)\dot{\bar{f}}(t)}{1+|f(t)|^2 }\,\ket{\omega}\!\bra{\omega}\,\dd t\: \right]\,.
\end{equation}
In the last step, we have recognized that this is now a unitary acting within the computational space $V(\bla_0)$, so we have expressed it in the corresponding basis, i.e., the computational basis. Indeed we consistently define $\ket{\omega}=\sum_{a=0}^{d-1} \omega_a \ket{a}$. We conclude that $U(T)=U_{(1)}$ has exactly the expression anticipated in Eq.~\eqref{eqs:our_gate_1and2qubits}, with the real number $\alpha_1$ given by %in Eq.~\eqref{eqs:alphas} with \textcolor{red}{[Discuss Stokes' theorem here:]}
\begin{equation}\label{eq:alpha1-derivation}
\alpha_1=\iu\int_0^T \frac{f(t)}{1+|f(t)|^2}\, \dot{\overline{f}}(t)\,\dd t \,.
\end{equation}
Notice that this integral can be interpreted as a line integral on the complex plane along the closed curve defined by %$f(t)$. 
$f$. We can now use Stokes' Theorem to transform this to a surface integral on the surface $S$ enclosed by the curve, which leads to an expression of the form as in Eq.~\eqref{eqs:alphas} with
\begin{equation}
    \mathcal{C}_1(z)=2{\left(\frac{1}{1+|z|^2}\right)}^2\,. \label{eq:C1}
\end{equation}

\subsubsection{Two-qudit unitaries}\label{sec:methods-2qudits}
Here, we derive the unitary $U(T)$ when the adiabatic loop is carried out on a pair of atoms interacting through the Hamiltonian~\eqref{eq:2-atom-hamiltonian}. In this case, 
the null subspace of Hamiltonian~\eqref{eq:2-atom-hamiltonian} is now $(d^2+1)$-dimensional and can be characterized by identifying the following linearly independent null eigenvectors. First, it is easy to see that the vector
\begin{equation}\label{eq:null-ev-2qudit_1}
    \ket{w^0}=\ket{+,-}-\ket{-,+}
\end{equation}
is in the null subspace of $H(\bla)$. Second, the vectors
\begin{equation}\label{eq:null-ev-2qudit_2}
    \ket{w^-_{ab}(\bla)}=\ket{e_a(\bla),e_b(\bla)}-\ket{e_b(\bla),e_a(\bla)}
\end{equation}
for all $0\le a<b\le d-1$, are orthogonal to $\ket{w^0}$, linearly independent and belong to the null subspace of $H(\bla)$. Third, the vectors
\begin{equation}\label{eq:null-ev-2qudit_3}
    \ket{w^+_{ab}(\bla)}=\ket{e_a(\bla),e_b(\bla)}+\ket{e_b(\bla),e_a(\bla)}+\frac{\overline{\Omega}_a \overline{\Omega}_b}{\Omega_d^2}(\ket{+,-}+\ket{-,+})
\end{equation}
for all $0\le a\le b\le d-1$, are orthogonal to $\ket{w^0}$ and to all $\ket{w^-_{ab}}$, linearly independent and belong to the null subspace of $H(\bla)$. Since we will see that the state vector $\ket{w^0}$ fully decouples from the others during the system's evolution, one can therefore treat the span of $\ket{w^-_{ab}}$ and $\ket{w^+_{ab}}$ as our effective computational subspace. We see that this is a $d^2$-dimensional subspace, which at the base point (that is characterized by $\Omega_a=0$ for all $0\le a\le d-1$) reduces to the tensor product of the two single-qudit computational subspaces.

Due to the interaction, it is now less trivial to fully characterize the remaining eigenvectors of the Hamiltonian~\eqref{eq:2-atom-hamiltonian}. However, we observe numerically that at finite interaction strength $W$ all the remaining eigenvectors have an energy separated from zero by a finite gap. We discuss this aspect in further detail in Appendix~\ref{sec:app_eig_hamiltonian}, where we show analytically that for large enough $W$ the gap can be lower bounded by $|\Om_d|/2$.

In summary, the null subspaces of the two-atom Hamiltonian in Eq.~\eqref{eq:2-atom-hamiltonian} are spanned by the basis vectors defined in Eqs.~\eqref{eq:null-ev-2qudit_1}--\eqref{eq:null-ev-2qudit_3}. As before, we collect in the vector $\boldsymbol{\lambda}$ the real parameters of the model, i.e., $\boldsymbol{\lambda}=(\mathrm{Re}\, \Omega_0,\dots, \mathrm{Re}\, \Omega_{d-1}, \mathrm{Im} \,\Omega_0,\dots, \mathrm{Im} \,\Omega_{d-1})$ and we assume the interaction strength $W$ and the parameter $\Omega_d$ to be fixed throughout the interaction. We observe that both $g$ and $A$ decompose into the direct sum of three blocks corresponding to the subspaces spanned by the vectors~\eqref{eq:null-ev-2qudit_1}, \eqref{eq:null-ev-2qudit_2} and~\eqref{eq:null-ev-2qudit_3} respectively (see Appendix~\ref{app:formulae} for a more detailed derivation of this). In other words
\begin{equation}
\begin{aligned}
    g&=g^0\oplus g^-\oplus g^+\,,\\
    A&=A^0\oplus A^-\oplus A^+\,.
\end{aligned}
\end{equation}
As we do not plan to encode any logical information in the subspace spanned by the vector~\eqref{eq:null-ev-2qudit_1}, we just observe that it remains uncoupled from the rest of the evolution and consequently ignore it. We instead focus on the dynamics in the subspaces spanned by the vectors~\eqref{eq:null-ev-2qudit_2} and~\eqref{eq:null-ev-2qudit_3}. Considering the block decomposition of $A$ into these two subspaces we have, up to a global phase,
\begin{equation}
U(T)=\mathcal{T} \exp \left[ -\int_0^T  \dot{\lambda}^\mu A_{\mu}(\boldsymbol{\lambda}(t)) \: \dd t \right]=U_+U_-
\end{equation}
where 
\begin{equation}
U_\pm=\mathcal{T} \exp \left[ -\int_0^T  \dot{\lambda}^\mu A^\pm_{\mu}(\boldsymbol{\lambda}(t)) \: \dd t \right]\,.
\end{equation}

Now we use 
Eqs.\ \eqref{eq:null-ev-2qudit_2}-\eqref{eq:null-ev-2qudit_3} to find explicit expressions for the previous unitaries. We start with $U^-$ by computing the matrix elements of the metric $g^-$ from the vectors~\eqref{eq:null-ev-2qudit_2}. From the definition, this can be written as
\begin{equation}
g^{-}_{(a,b),(k,l)}(\bla):=\braket{w_{ab}^-(\bla)|w^-_{kl}(\bla)}=2\left(g_{ak}(\bla)\,g_{bl}(\bla)-g_{bk}(\bla)\,g_{al}(\bla) \right)
\end{equation}
where $0\le a<b\le d-1$ and $0\le k<l\le d-1$, and with $g_{ab}(\bla)$ as in Eqs.~\eqref{eq:g-1qudit}. From here, one can analytically derive the matrix elements of the inverse of the metric as
\begin{equation}
g^{-\,(a,b),(k,l)}(\bla)=\frac{1}{8} \left( g^{ak}(\bla)\,g^{bl}(\bla)-g^{bk}(\bla)\,g^{al}(\bla)\right)\,
\end{equation}
with $g^{ab}(\bla)$ as in Eqs.~\eqref{eq:gi-1qudit}. We further have
\begin{equation}
\begin{aligned}
    A^{-}_{(k,l),(a,b),c}(\bla)&:=\braket{w^-_{kl}(\boldsymbol{\lambda})|\partial_c w^-_{ab}(\boldsymbol{\lambda})} \\
    &=2\left( A_{kac}(\bla) \, g_{lb}(\bla) + A_{lbc}(\bla) \, g_{ka}(\bla) - A_{kbc}(\bla) \, g_{la}(\bla) - A_{lac}(\bla) \, g_{kb}(\bla)\right)\,,
\end{aligned}
\end{equation}
where $A_{abc}(\bla)$ is defined as in equations~\eqref{eq:A-1qudit}. Now take the parameter loop defined by $\Omega_a(t)=|\Omega_d| \, f(t)  \, \omega_a$ for $0\le a\le d-1$, $f$ is a complex function such that $f(0)=f(T)=0$ and $\sum_a |\omega_a|^2=1$. Combining the results above, we arrive at
\begin{equation}
    \dot{\lambda}^\mu A^{- \, (a,b)}{}_{(k,l),\mu}(t)  = \frac{f(t)\dot{\overline{f}}(t)}{2+2|f(t)|^2}\left(\delta_{ak} \, \omega_b \overline{\omega}_l + \omega_a \overline{\omega}_k \, \delta_{bl} - \delta_{al} \, \omega_b \overline{\omega}_k - \omega_a \overline{\omega}_l \, \delta_{bk} \right)\,,
\end{equation}
where $0\le a<b\le d-1$ and $0\le k<l\le d-1$, and $\de_{ab}$ is the Kroenecker delta. Equivalently, in matrix notation
\begin{equation}
\dot{\lambda}^\mu A^-_\mu(t)=\frac{f(t)\dot{\overline{f}}(t)}{1+|f(t)|^2} \, \left(1\otimes\ket{\om}\!\!\bra{\om}+\ket{\om}\!\!\bra{\om}\otimes 1\right) \left(\frac{1-S}{2}\right)\,,
\end{equation}
where $S:\ket{a,b}\mapsto\ket{b,a}$ is the two-qudit SWAP operator and $\ket{\omega}=\sum_{a=0}^{d-1} \omega_a \ket{a}$ as before. Since one can see that these matrices commute at different times, we conclude that
\begin{equation}\label{eq:time-ordered-int-2qudit_-}
        U_-=\mathcal{T} \exp \left[ -\int_0^T  \dot{\lambda}^\mu A^-_{\mu}(\boldsymbol{\lambda}(t)) \: \dd t \right]=\exp\left[\iu\,\alpha_1 \left(1\otimes\ket{\om}\!\!\bra{\om}+\ket{\om}\!\!\bra{\om}\otimes 1 \right) \left(\frac{1-S}2\right)\right]
\end{equation}
where $\alpha_1$ is as given in Eq.~\eqref{eq:alpha1-derivation}.

Now one can also consider the block corresponding to the vectors~\eqref{eq:null-ev-2qudit_3} and follow the same steps to compute $U_+$. That is, one can use Eqs.~\eqref{eq:null-ev-2qudit_3} to derive analytical expressions for
\begin{equation}
g^{+}_{(a,b),(k,l)}(\bla):=\braket{w_{ab}^+(\bla)|w^+_{kl}(\bla)}
\end{equation}
where $0\le a\le b\le d-1$ and $0\le k\le l\le d-1$. This allows one to analytically compute the matrix elements $g^{+\,(a,b),(k,l)}(\bla)$ of the inverse metric. The expressions of both these quantities are somewhat cumbersome, so we defer them to Appendix~\ref{app:formulae} for the interested readers. Furthermore, one can also compute
\begin{equation}
A^{+}_{(k,l),(a,b),c}(\bla):=\braket{w^+_{kl}(\boldsymbol{\lambda})|\partial_c w^+_{ab}(\boldsymbol{\lambda})}
\end{equation}
analytically. Again, this is a cumbersome expression that can be found in Appendix~\ref{app:formulae} in full.

Now we take the same path as above, i.e., the parameter loop defined by $\Omega_a(t)= |\Omega_d| \, f(t) \, \omega_a$ for $0\le a\le d-1$, where $f(t)$ is a complex function such that $f(0)=f(T)=0$ and $\sum_a |\omega_a|^2=1$. Putting this all together with the expressions mentioned above, one can derive analytical expressions for $\dot{\lambda}^\mu A^{+ \, (k,l)}{}_{(a,b),\mu}(\bla)$ for $0\le a\le b\le d-1$ and $0\le k\le l\le d-1$. In matrix notation this has the simple form
\begin{equation}
\dot{\lambda}^\mu A^+_\mu(t)=B_1(t)\left(1\otimes\ket{\om}\!\!\bra{\om}+\ket{\om}\!\!\bra{\om}\otimes 1\right)\left(\frac{1+S}{2}\right)+  B_2(t)(\ket{\om}\!\!\bra{\om}\otimes\ket{\om}\!\!\bra{\om})\,,
\end{equation}
where $B_1(t)$ and $B_2(t)$ are given by
\begin{align}%\label{eq:B1_B2}
    B_1(t)&=\frac{f(t)\dot{\bar{f}}(t)}{1+|f(t)|^2 }, 
    \\
    B_2(t)&=\frac{ 3 \dot{f}(t)\overline{f}(t) |f(t)|^4 +f(t) \dot{\overline{f}}(t) |f(t)|^2  (4+ |f(t)|^2 )} {(1+|f(t)|^2) \, (1+2|f(t)|^4)}\,.
\end{align}
Again, one can see that these matrices commute at different times and, therefore,
\begin{multline}\label{eq:time-ordered-int-2qudit_+}
 U_+=\mathcal{T} \exp \left[ -\int_0^T  \dot{\lambda}^\mu A^+_{\mu}(\boldsymbol{\lambda}(t)) \: dt \right]\\
=\exp\left[\iu\,\alpha_1\left(1\otimes\ket{\om}\!\!\bra{\om}+\ket{\om}\!\!\bra{\om}\otimes 1\right)\left(\frac{1+S}2\right)\right]\exp[\iu\,\alpha_2\ket{\om}\!\!\bra{\om}\otimes\ket{\om}\!\!\bra{\om} ]\,, 
\end{multline}
where $\alpha_1$ is the same as in~\eqref{eq:alpha1-derivation} and $\alpha_2$ is the real number given by the integral of $\iu B_2(t)$ over $[0,T]$. As before we can view this integral as a line integral and convert it, using Stokes' Theorem, to a surface integral on the complex plane. This leads to an expression of the form of Eqs.~\eqref{eqs:alphas} with
\begin{equation}
    \mathcal{C}_2(z)=\frac{4 |z|^2 (4 - |z|^2 -2 |z|^4 -6 |z|^6)}{(1+|z|^2)^2 (1+2|z|^4)^2}. \label{eq:C2}
\end{equation}
Finally, combining Eqs.~\eqref{eq:time-ordered-int-2qudit_-} and~\eqref{eq:time-ordered-int-2qudit_+} one arrives at the desired result: the full gate produced by the given parameter loop is
\begin{align}\label{eq:U2}
    U(T)&=U_+U_- = \exp\left[\iu\alpha_1\left(1\otimes\ket{\om}\!\!\bra{\om}+\ket{\om}\!\!\bra{\om}\otimes 1\right)\right]\cdot\exp\big[\iu\alpha_2(\ket{\omega}\!\otimes\! \ket{\omega})\,(\bra{\omega}\!\otimes\! \bra{\omega})\big]. %\exp\left[\iu\alpha_1 (1\otimes \omega \,\omega^\dag + \omega \,\omega^\dag \otimes1) \right] \, \exp[\iu \alpha_2\: (\omega \otimes \omega) \,(\omega \otimes \omega)^\dag ]\,.
\end{align}
We conclude that $U(T)=U_{(2)}$ has the form anticipated in Eq.~\eqref{eqs:our_gate_1and2qubits} with $\alpha_1$ and $\alpha_2$ as given in Eqs.~\eqref{eqs:alphas} and with $\mathcal{C}_1(t)$ and $\mathcal{C}_2(t)$ given by Eqs.~\eqref{eq:C1} and~\eqref{eq:C2}.

\subsection{Error analysis}
 \label{sec:methods-errors}
In this section we present analytical and numerical investigations of the dependence and robustness of our  holonomic gates presented in Eq.~(\ref{eqs:our_gate_1and2qubits}) to various types of errors.
%We can identify three main sources of error that affect the implementation of our gate. 
Most importantly, we explain more in depth in Sec.~\ref{sec:methods-errors-coherent} how our holonomic quantum gates exhibit a geometric robustness to coherent errors. Besides coherent errors, the implementation of our gate will always be affected by an adiabatic error stemming from the approximated adiabatic evolution because of the finite protocol time.  Also decoherence errors are likely to be present in the experimental implementation due to the finite life time of the atomic energy levels. We present a numerical analysis of the influence of the adiabatic error and decoherence under a simple model and their combination in Sec.~\ref{sec:methods-errors-decoherence}. Additionally, in Sec.~\ref{sec:incoherent} we comment on how the robustness to coherent errors of our gates can give rise to a robustness from specific families of environment induced noise.

In all our investigations, as a figure of merit for the implementation of our gates we consider~\cite{fromonteil_protocols_2023} the gate fidelity %$\mathcal F$ 
\begin{equation}
     \mathcal F = \frac{D +\left|\tr{\tilde U U^\dagger}\right|^2}{D(D+1)},
     \label{equ:gate_fidelity}
\end{equation}
between the unitary $U$ we wish to implement and the noisy unitary $\tilde U$ implemented under the influence of errors and noise,
with $D$ being the dimension of $U$ (or equivalently the dimension of the computation subspace). This gate fidelity is related to common
measures for the quality of implementations of quantum gates, such
as the average gate fidelity~\cite{BenchmarkingReview}.

\subsubsection{Geometrical protection from coherent errors}\label{sec:methods-errors-coherent}

 Coherent errors can be seen as (slight) deviations from the original path in the parameter space which result in effectively evolving along a deformed loop in the parameter manifold. 

As outlined above, the real and imaginary parts of the complex transition amplitudes are the tunable parameters of our system and we chose them to be of the form $\mathbf{\Omega}(t) = \bm{\omega} f(t)$. In the following analysis we assume that the noisy loop in parameter space can also be expressed as $\mathbf{\tilde \Omega} (t)=\bm{\omega}\tilde f(t)$. This corresponds to noise which acts with the same time-dependence on all transition amplitudes and does not affect the constant unit vector $\bm{\omega}$. More specifically,  we consider a deviation of the complex loop $f(t)$ to be of the form
\begin{equation}
 f(t)\mapsto\tilde f(t) = \bigl(1 +\epsilon(t)\bigr)f(t)\ee^{i \phi(t)},
 \label{eq:coh_error_model_loop}
\end{equation}
 where $ \epsilon(t),\phi(t) \in \mathbb R$. Of course generally we will have that $\max_t|\epsilon(t)| \ll 1$. The altered loop $\tilde f(t)$ will in general enclose a different area $\tilde S$ than the area $S$ enclosed by $f(t)$. This means that the surface integrals in Eq.~(\ref{eqs:alphas}), which determine the unitary $U$, have different values $\tilde \alpha_1, \tilde \alpha_2$ and hence a different unitary $\tilde U$ gets implemented. From
\begin{equation}
\label{eq:coherr_gatefid_step1}
    \begin{aligned}
    \Bigl|\tr{\tilde U_{(1)} U_{(1)}^\dagger} \Bigr|^2 &= \Bigl|\tr{\ee^{i (\tilde \alpha_1 -\alpha_1) \ket{\omega}\bra{\omega}}} \Bigr|^2,  \\
        \Bigl|\tr{\tilde U_{(2)} U_{(2)}^\dagger} \Bigr|^2 &= \Bigl|\tr{\ee^{i (\tilde \alpha_1-\alpha_1)(\id \otimes \ket{\omega}\bra{\omega}+ \ket{\omega}\bra{\omega}\otimes \id)}\ee^{i (\tilde \alpha_2 -\alpha_2) (\ket{\omega}\otimes \ket{\omega})(\bra{\omega}\otimes \bra{\omega})}} \Bigr|^2, 
    \end{aligned}
\end{equation}
we see that the gate fidelity $\mathcal F$ (\ref{equ:gate_fidelity}) then depends on the differences between the two surface integrals $\delta \alpha_j := \tilde \alpha_j - \alpha_j$, for $j=1,2$, with
\begin{equation}
\begin{aligned}
     |\delta \alpha_j| &= \Bigl|\int_{\tilde S}  \mathrm d S \; \mathcal C_j(z) - \int_{S} \mathrm d S \; \mathcal C_j(z) \Bigr|
    &\leq  \int_{\delta S}  \mathrm d S \; |\mathcal C_j(z)| .
\end{aligned}
\end{equation}
 Here, we have defined $\delta S = (\tilde S \cup S) \backslash (\tilde S \cap S)$ to be all the points which are exclusively either in $S$ or in $\tilde S$. The difference between the two phases is thus the surface integral over the area $\delta S$ and it hence depends on the value of the function $\mathcal C_i$ on $\delta S$.
\begin{figure}
    \centering
        \includegraphics[width=0.55\textwidth]{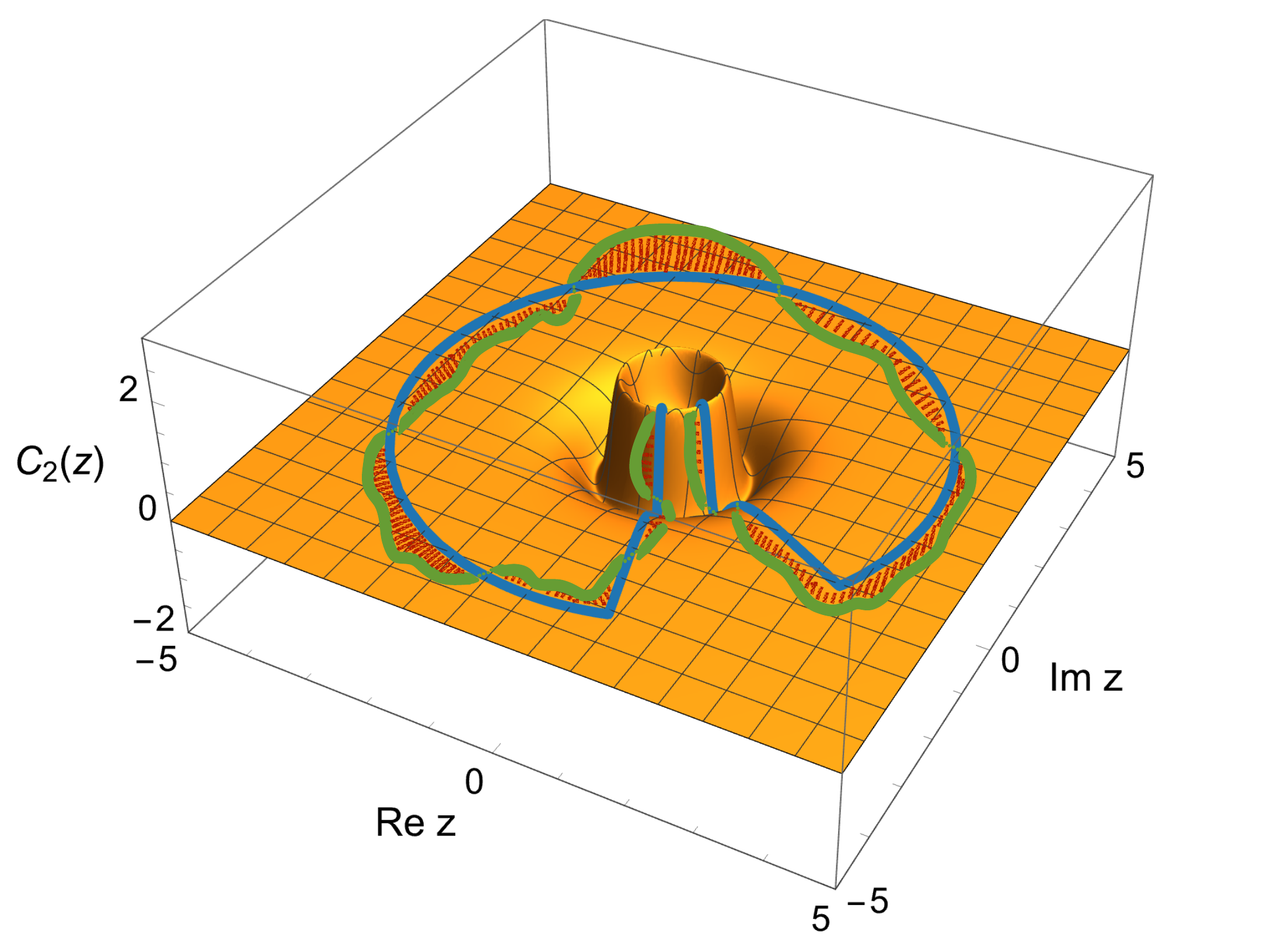}
\caption{The difference surface (red) between the original loop $f(t)$ (blue) and the noisy loop $\tilde f(t)$ (green) determines the gate fidelity via surface integrals. The decay of the integrands can be used to diminish the value of this surface integral. This can be achieved by performing the loop mostly in regions of vanishingly small integrand  as shown in this figure for $\mathcal C_2(z)$. }
    \label{fig:deltaS_C1C2_integrandfunctions}
\end{figure}
 As we have already seen in Fig.~\ref{fig:general_loop_integrands-b}, the functions $\mathcal C_1(z)$ and $\mathcal C_2(z)$ decay rapidly. It is precisely this decay which grants our gates a robustness to coherent errors. This is because we can choose the original loop $f(t)$ to run mostly through regions where the $\mathcal C_i(z)$'s are very small and consequently its deformations in these regions will have a small impact on the overall gate fidelity through smaller values of $\delta \alpha_1, \delta \alpha_2$ (Fig.~\ref{fig:deltaS_C1C2_integrandfunctions}). We will make this more concrete below by choosing a specific loop $f(t)$. At this point, we point out a conceptual connection between our holonomic implementation and the geometric perspective on fault-tolerant gates discussed in Ref.~\cite{gottesman2017fibre}. In that work, certain fault-tolerant gate constructions on quantum error-correcting codes are interpreted geometrically, with logical operations corresponding to loops enclosing regions of zero curvature in a manifold associated with the code space.
 
Although our setting is different, the robustness of our holonomic gates against certain coherent control errors can likewise be understood in geometric terms, as arising from properties of the path followed in parameter space. In our case, however, the “curvature” of the manifold—characterized by the values of $\mathcal C_i(z)$ along the loop—is only approximately zero in regions of large $|z|$, and the loop cannot be placed entirely within regions of vanishing curvature. We emphasize that our proposal does not constitute fault tolerance in the usual sense: the computational subspaces we consider are not error-correcting code spaces, and our gates do not by themselves prevent the propagation of generic errors. Rather, our results demonstrate a form of hardware-level robustness (error mitigation) whose origin is geometric, in a way that is conceptually analogous to the geometric interpretation of fault-tolerant gates in Ref.~\cite{gottesman2017fibre}.
 
\begin{figure}
        \centering
        \includegraphics[width=0.4\textwidth]{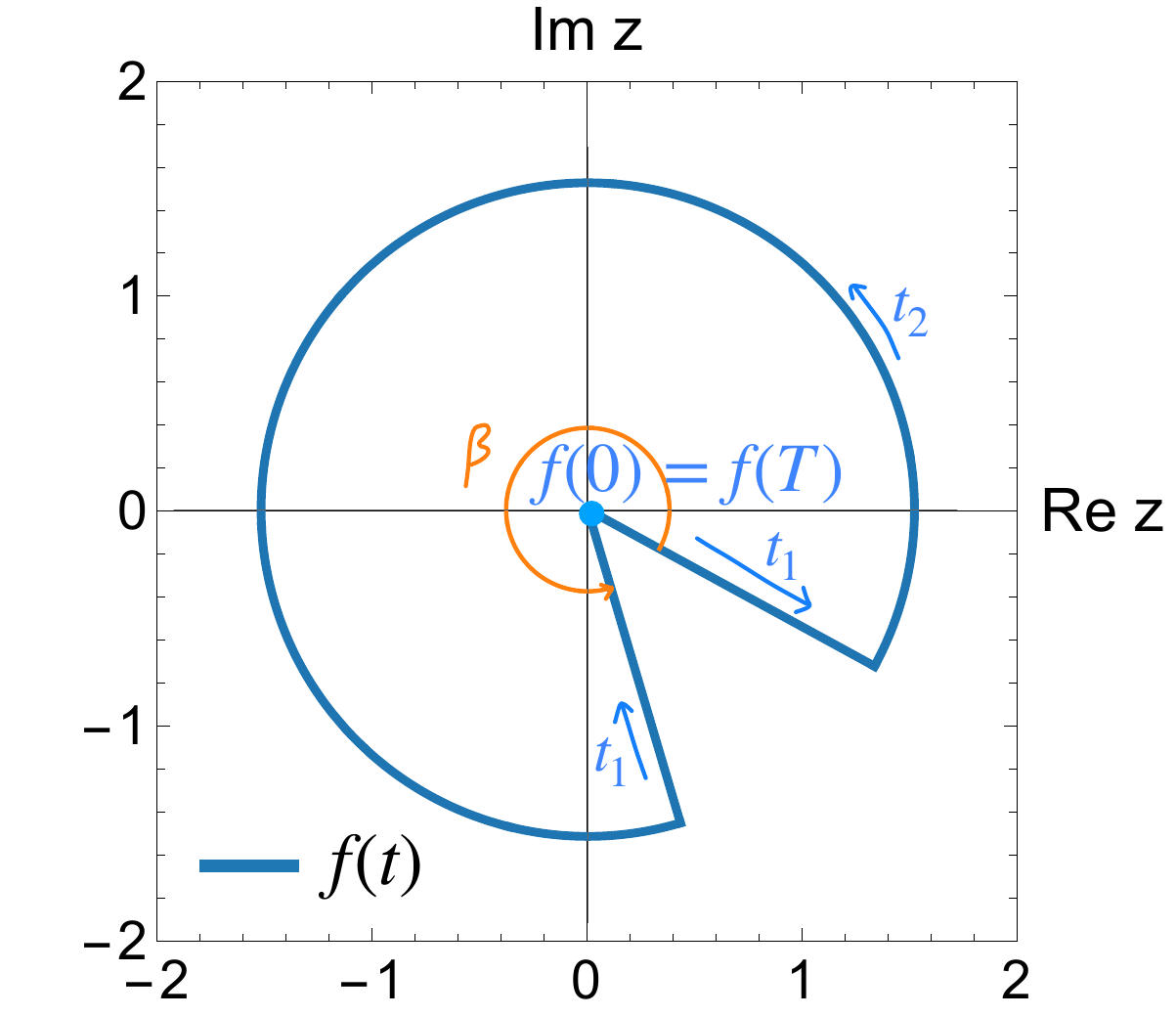}
\caption{A particular choice for the loop $f(t)$ (Eq.~\eqref{eq:simple-loop-def}) in the complex plane implementing our phase gate. We linearly parametrise every of the three segments and have a total protocol time $T = 2t_1 + t_2$. At a constant radius $R$ we can reach any phase $\alpha_2$ for our gate by appropriate choice of the angle $\beta$.}
    \label{fig:pacman_loop}
\end{figure} 
In the following and for the rest of this section, we focus on the two-qudit gate $U_{(2)}$ with $D=d^2$ and consider a particular implementing loop $f(t)$ with (see also Fig.~\ref{fig:pacman_loop}) 
\begin{equation}
    \begin{aligned}
f(t)&=\begin{cases}
        R \, \frac{t}{t_1}, & t \in[0,t_1),\\
        R \, \ee^{\iu \beta \frac{t-t_1}{t_2}},&t\in[t_1,t_1+t_2),\\
        R \,\frac{2t_1+t_2-t}{t_1}\ee^{\iu \beta},&t\in[t_1+t_2,2t_1+t_2]\je{.}
    \end{cases}
\end{aligned}
\label{eq:simple-loop-def}
\end{equation}
 The loop is linearly parametrized in time with $t_1$ denoting the time 
 to reach the outer circle of radius $R$ from the origin and $t_2$ being the time to wrap around the origin at constant radius up to an angle $\beta$. Hence, the total protocol time is $T = 2t_1 + t_2$. Since the expressions for  the phases $\alpha_i$, $i=1,2$ in our gates from Eqs.~(\ref{eqs:alphas}) can then be written as
\begin{equation}
    \alpha_i = \beta \cdot \int_0^R \mathrm d r \, \mathcal C_i(r) \cdot r,
\end{equation}
 we see that at a fixed radius $R$ we can reach any phase $\alpha_2$ by a suitable choice of $\beta$. 

Now, we consider a constant offset error $\epsilon(t) \equiv \epsilon$ and no phase error which corresponds to $\tilde f(t) = (1 + \epsilon)f(t)$. This is an interesting example where we can  analytically bound the gate fidelity $\mathcal F$ to leading order in $\epsilon$. The result can be interpreted as a lower bound on the fidelity for all time-dependent multiplicative errors $\epsilon(t)$ with $\epsilon = \max_t|\epsilon(t)|$. Expanding the exponential in Eq.~\eqref{eq:coherr_gatefid_step1} up to second order in the error we get for the gate fidelity
\begin{equation}
    \begin{aligned}
           \mathcal F  &= 1 - \frac{2(d-1)}{d^2+1}(\delta \alpha_1)^2 - \frac{d^2-1}{d^2(d^2+1)} (\delta \alpha_2)^2 - \frac{4(d-1)}{d(d^2+1)} \delta \alpha_1 \delta \alpha_2 + \mathcal{O}((\delta \alpha)^3).
    \end{aligned}
\end{equation}
For the phase differences, we find
\begin{equation}
\begin{aligned}
    \delta \alpha_2 &=  \beta \int_{R}^{R(1+ \epsilon)} \mathrm d r \; r \; \mathcal C_2(r)  = \frac{\beta 4R^4(4-R^2-2R^4-6R^6)}{(1+R^2+2R^4+2R^6)^2} \epsilon + \mathcal{O}(\epsilon^2), \\
    \delta \alpha_1 &= \beta \int_{R}^{R(1+\epsilon)} \mathrm d r \; r \; \mathcal C_1(r)  =  \frac{\beta 2 R^2}{(1+R^2)^2} \epsilon+ \mathcal{O}(\epsilon^2), 
\end{aligned}
\end{equation}
and consequently 
\begin{equation}
     \mathcal F  = 1 - c(R) \beta^2 \epsilon^2 + \mathcal{O}(\epsilon^3),
\end{equation}
where the coefficient $c(R)$ scales asymptotically like $c(R) \in \mathcal{O}({R}^{-4})$.

\begin{figure}
\centering
\includegraphics[width=0.6\textwidth]{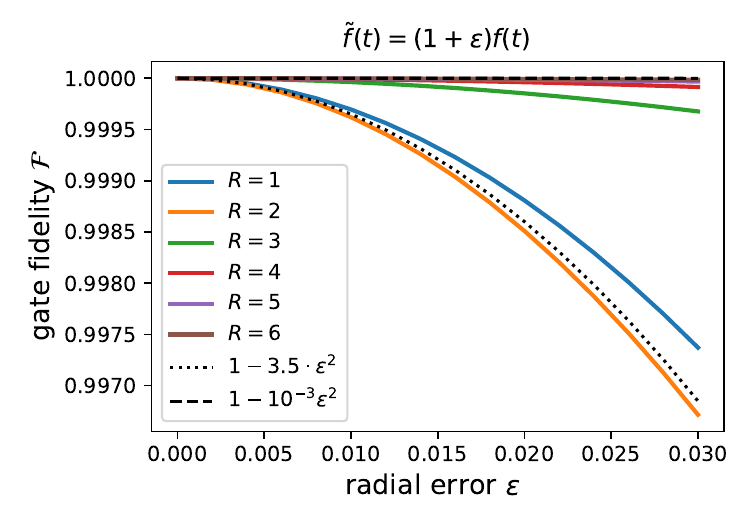}
\caption{For the case $d=2$ we computed the gate fidelity for a constant multiplicative error $\epsilon$. The analytically found scaling $1-C\cdot\epsilon^{2}$ is apparent. We see as well how the coefficient in the leading-order error term can be decreased by choosing higher values of $R$ for $R > 3$.
    }
    \label{fig:numerics_coh_error}
\end{figure}

We can arbitrarily decrease the coefficient $c(R)$ in the leading order term of the gate fidelity by choosing a higher radius $R$ in the complex loop $f(t)$. Due to the rapid decay of $c(R)$ already slightly increasing $R$ makes our gate significantly more robust. This is a direct manifestation of the geometrical robustness of our gates to coherent errors. Notice that a quadratic scaling $1-\mathcal{F}=\mathcal{O}(\epsilon^2)$ of the fidelity with respect to the error magnitude was already observed in reference~\cite{fromonteil_protocols_2023} for a variety of gate protocols in Rydberg atoms. In these cases, however, the leading term coefficient was a fixed property of the protocol and could not be reduced like in our holonomic model. 
We also investigated the gate fidelity numerically for $d=2$ and the noiseless gate being the controlled-$Z$ (see Fig.~\ref{fig:numerics_coh_error}). It can be readily seen how the susceptibility to coherent errors is increased by choosing loops with a higher radius $R$. 
Additionally, we observe that the coefficient in the leading $\epsilon^2$-term of the gate fidelity is reduced by multiple orders of magnitude between values $R=2,3,4$ for the radius of the original loop, as expected from the theoretical analysis above. 

Finally, we have to keep in mind that the deviations of the path around the origin of the complex plane will always induce a non-vanishing contribution to the coherent error as the functions $C_1, C_2$ exhibit a large variation there. In our proposed gates the path $f(t)$ has to have its base point at the origin. Thus, there is a priori no possibility to choose a path which does not cross at all through regions of ``large curvature''. For our loop parameterization as in Eqs.~(\ref{eq:simple-loop-def}) and the error model Eqs.~(\ref{eq:coh_error_model_loop}) this means that the phase error $\phi(t)$ from the first and third part of the loop will induce a non-vanishing contribution to the error due to fluctuations close to the origin. The phase errors in the second, angular part of the loop however will only change the parametrization of the loop and thus not alter its shape. It is an interesting open direction to investigate if one can devise practical holonomic gates generated by paths which just encircle regions of non-trivial curvature but do not cross them.

\subsubsection{Adiabatic approximation and decoherence errors}
\label{sec:methods-errors-decoherence}
\begin{figure}
    \centering
        \begin{subfigure}[b]{0.48\textwidth}
        \centering
    \includegraphics[width=0.96\textwidth]{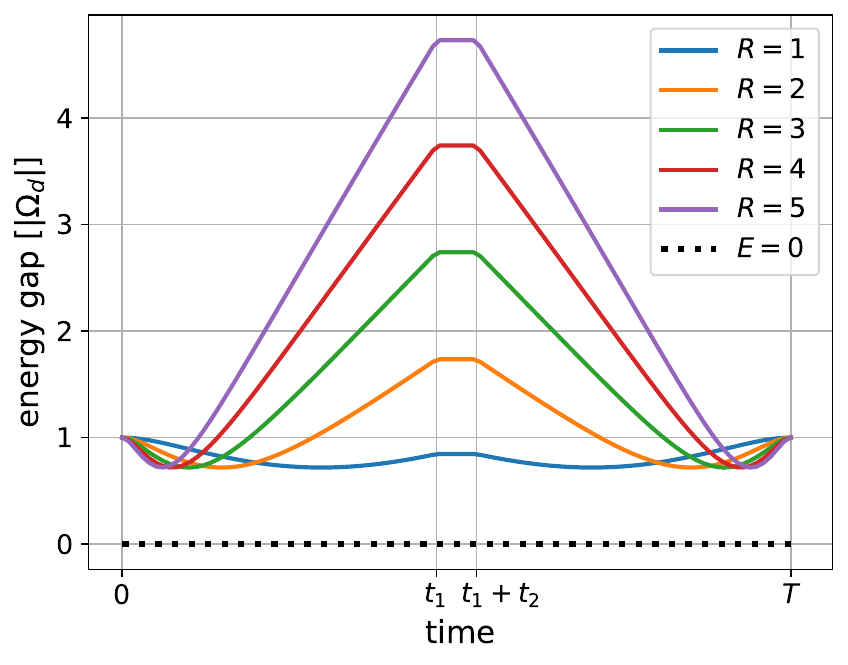}
        \caption{}
        \label{fig:gap_hamil-a}
    \end{subfigure}
    \begin{subfigure}[b]{0.48\textwidth}
        \centering
        \includegraphics[width=\textwidth]{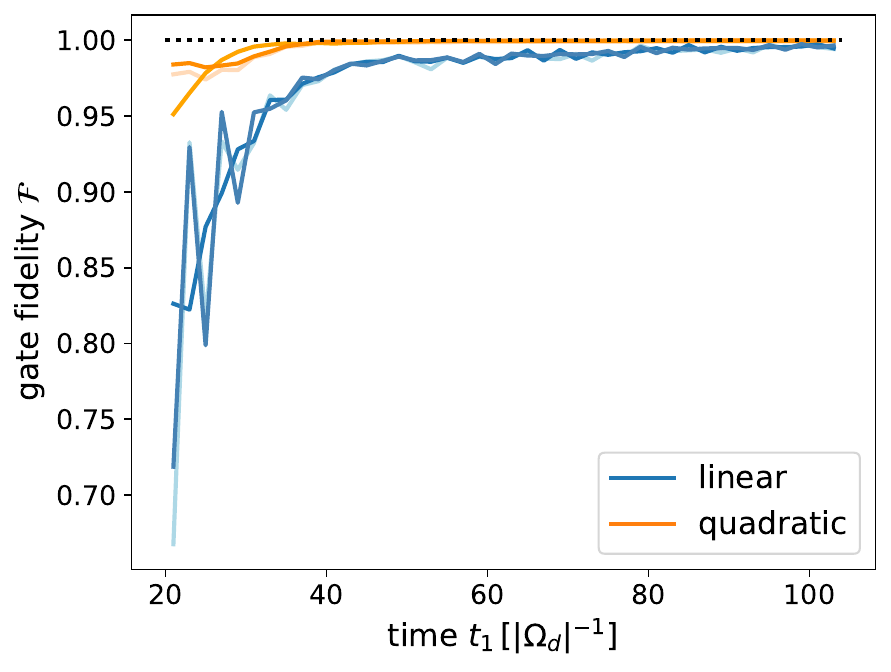}
        \caption{}
        \label{fig:gap_hamil-b}
    \end{subfigure}
    \caption{a) The closest eigenvalue of the Hamiltonian to the zero-energy computation subspace when evolving along the loop $f(t)$. The smallest value is in the ``outgoing'' and ``incoming'' part of the loop, and it is independent for the investigated range of radii $R$. b) Changing the linear parametrization in time to a quadratic parametrization ($\frac{t}{t_1} %\rightarrow 
    \mapsto
    (\frac{t}{t_1})^2)$ means that around the origin of the complex plane we move slower which is exactly where the gap is smallest. A fixed gate fidelity is reached at already much shorter times $t_1$. The three different shades of the blue/orange lines correspond to different times $t_2 = $ 10/20/30 $[|\Omega_d|^{-1}]$.}
\end{figure}

The adiabatic error and the decoherence can be seen as two antagonists since the former asks for as long as possible protocol times while the latter has smaller impact with faster protocol times. We numerically analyze the impact of the adiabatic and decoherence errors for the controlled-$Z$ entangling gate which corresponds to $d=2$, $\alpha_2 = \pi$, and $\bm{\omega} = (0,1)^{\text{T}}$, up to single-qubit rotations. Our choice of parametrising the loop implies that the time parameter is expressed in units of $|\Omega_d|^{-1}$, the inverse fixed transition amplitude. In the experimental implementation this would corresponds to the inverse Rabi frequency and the level $\ket{d}$ to some Rydberg state (see Appendix~\ref{sec:app_Hphysical}). It is this Rydberg state that is most strongly affected by decoherence \cite{saffmanAnalysisQuantumLogic2005}. In all numerical simulations we set the maximal interaction strength to be  $W = 10 \,[|\Omega_d|]$. This is because the dependence on $W$ of the energy gap in the Hamiltonian, which determines the adiabatic error, saturates at around $W \approx 10 \,[|\Omega_d|]$, as we numerically observe (see Fig.~\ref{fig:gaps_Rydinter_strength} in Appendix~\ref{sec:app_eig_hamiltonian}). Consequently, increasing the interaction strength further does not lead to a decrease of the adiabatic error. 
We include decoherence of the state $\ket{d}$ to the simulation by modelling its finite life-time with a non-hermitian decay term which is added to both single-atom states \cite{jakschFastQuantumGates2000,jandura_time-optimal_2022}. More concretely, the Hamiltonian $H_0(\bm{\lambda})$ in Eq.~\eqref{eq:2-atom-hamiltonian} is extended by the term $\frac{\Gamma}{2i} \ket{d}\bra{d} $ with $\Gamma$ being the decay rate (inverse life-time) of the level $\ket{d}$. In a sufficiently controllable experimental setup $\Gamma \ll |\Omega_d|$ and we set here $\Gamma= 10^{-4}, 10^{-3} \,[|\Omega_d|]$ \cite{saffmanAnalysisQuantumLogic2005}.  Consequently, to extract the effectively implemented unitary after time-evolving for some finite protocol time $T$ under the Hamiltonian subject to decoherence of the state $\ket{d}$, we numerically solve the time-dependent Schrödinger equation. We are interested in the total protocol time $T$ for which the combination of decoherence and adiabatic error has the smallest impact on the gate fidelity $\mathcal F$.
\begin{figure}
    \centering
    \includegraphics[width=0.6\textwidth]{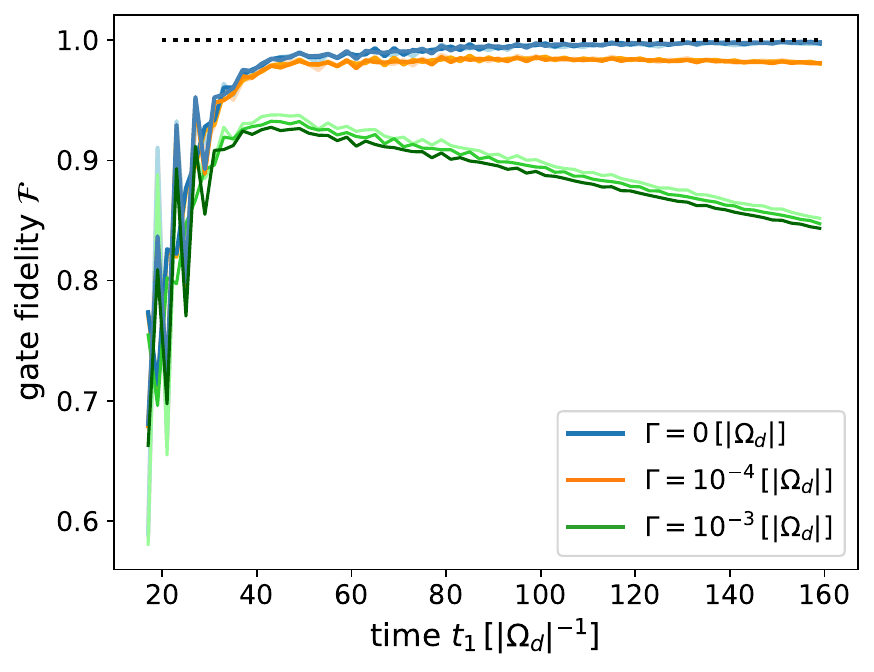}
    \caption{We compute the gate fidelity $\mathcal F$ between the target controlled-$Z$ gate and the effectively implemented gate resulting from a near-adiabatic evolution along $f(t)$ with $R=5$ for different partial protocol times $t_1$ and $t_2$. The gate is determined by numerically solving the time-dependent Schrödinger equation. A non-Hermitian term proportional to the decay rate $\Gamma$ of the state vector $\ket{d}$, \atoms{usually some Rydberg level},  models decoherence \cite{jakschFastQuantumGates2000,jandura_time-optimal_2022}.  The three different shades of the blue/orange/green lines in the plot correspond to the different times $t_2 = $  10/20/30 $[|\Omega_d|^{-1}]$, respectively. We see that the error is dominated by the value of $t_1$ compared to $t_2$ because the variation between the different shades of a color is very small. This is because on the loop segments governed by $t_1$ (outgoing and incoming part) the gap of the Hamiltonian is much smaller. Overall, at small protocol times the adiabatic error is dominant but at later times the decoherence becomes more significant. Therefore, in the presence of decoherence ($\Gamma \neq 0$), there is a sweet-spot of the protocol time corresponding to a maximum of the gate fidelity.}
    \label{fig:ad_decoh_plot}
\end{figure}
Fig.~\ref{fig:ad_decoh_plot} shows for our controlled-$Z$ phase gate with radius $R=5$ the gate fidelity as a function of different protocol times $T$. We find that the adiabatic error is dominated by the effect of the outgoing and incoming time $t_1$ compared to the effect of the time $t_2$. This comes as no surprise when taking into account the size of the energy gap along the path (Fig.~\ref{fig:gap_hamil-a}). Furthermore, we can observe how for small protocol times the adiabatic error is dominant but at longer times the decoherence becomes more significant. For a decay rate $\Gamma = 10^{-4} \,[|\Omega_d |]$ and a total protocol time $T = 168 \,[|\Omega_d |^{-1}]  $ we achieve a gate fidelity of $\mathcal F = 98.80\% $.

Finally, we want to mention how the performance of our gates can be improved. The adiabatic error depends on the size of the energy gap between the eigenvalue zero  (the associated eigenspace is the computational subspace) and the closest eigenvalue. To keep the adiabatic error constant, we have to pass slower in regions of the parameter space where this gap is small. Therefore, we can decrease the protocol time needed by altering the parametrization of the loop $f$
according to the energy gap without penalty of a higher adiabatic error. Or equivalently, we can increase the gate fidelity achieved at a fixed protocol time. For our phase gate, Fig.~\ref{fig:gap_hamil-a} shows the gap of the Hamiltonian along the loop $ \mathbf \Omega(t)$ in parameter space.
The gap is smallest at the origin and increases monotonically towards the outer circle of the loop. There, the gap is constant as expected since the eigenvalues of the Hamiltonian depend only on the absolute values $|\Omega_a(t)|$ (for more discussion on the gap see Appendix~\ref{sec:app_eig_hamiltonian}). This suggests that a non-linear parametrization of the ``outgoing'' and ``incoming'' part of the loop is better than a linear one in terms of the achieved gate fidelity. The effect of just altering the parametrization along the outgoing and incoming part of the loop from linear to quadratic is already significant (see Fig.~\ref{fig:gap_hamil-b}). Using such an improved parametrization we numerically find that we can 
achieve a gate fidelity of $\mathcal F = 99.38\% $ within a total protocol time $T= 66 \,[|\Omega_d|^{-1}]$ with $R=5$ and $\Gamma=10^{-4} \, [|\Omega_d|] $. In a similar spirit, also the shape of the loop can be optimized to give smaller adiabatic errors using advanced optimized control techniques~\cite{jandura_time-optimal_2022,PhysRevResearch.5.033052}.
%[more citations]}. 
This, however, lies outside the scope of this work and constitutes a natural next step to further improve the performance of our gates. 

\subsubsection{Robustness to incoherent errors from dynamical disorder}\label{sec:incoherent}

The main strength of the analysis above is that it shows the inherent robustness of the scheme to coherent errors in the classical control. However, it would also be important to see to what extent holonomic schemes may be able to tolerate incoherent errors that resemble environment induced decoherence. This is particularly important in the light of potential implementations in Rydberg atom platforms~\cite{bernien_probing_2017,Levine19,Controlling,PhysRevLett.128.113602, cong_hardware-efficient_2022,bluvstein_quantum_2022,BlochArray,bluvstein_quantum_2024}.
An idea to understand this is to observe that some incoherent errors can be written as convex combinations of coherent errors, like the ones that give rise to Eq.~(\ref{eq:delta-alpha}). 

To be concrete, let us consider a functional dependence in time of the control parameters of the form
\begin{equation}
\Omega_a(t) = \Omega_a^0(t)+\gamma\xi_a(t), \label{eq:omega+xi}
\end{equation}
where $\xi_a$ is a stochastic fluctuating noise term in addition to the desired and anticipated time-controlled term $\Omega_a^0$, for $a=0,\dots, d-1$ and $\gamma$ is a real parameter determining the noise strength. To make some progress, let us assume that $\xi_a(t)$ are Gaussian stochastic processes, with vanishing mean and two-time correlations functions given by
\begin{equation}
    \mathbb{E}[\xi_a(t)]=0\,, \hspace{10mm} \mathbb{E}[\xi_a(t)\xi_b(s)]=\delta_{ab} \, D_{a}(t,s), \label{eq:main-correlation-function}
\end{equation}
where $\mathbb{E}$ indicates the mean over the stochastic realisations of the noise process. For Gaussian processes, these two quantities fully determine the process. In particular, if $D_a(t,s)$ is sufficiently regular, we can assume $\xi_a(t)$ to be smooth functions. By adding this noise to the Schr\"odinger equation of the system, the time evolution of the quantum state is then given by  $\ket{\psi(t,\xi)}$, where the state now depends explicitly on the realisation of the noise process $\xi=(\xi_0,\dots, \xi_{d-1})$. The key insight of our previous analysis of coherent errors is that the evolved state should only weakly depend on this noise. 

Considering now the probabilistic mixture over noise realisations we have the mixed state 
\begin{equation}
    \rho(t)=\mathbb{E} \left[ \ket{\psi(t,\xi)}\!\bra{\psi(t,\xi)} \right], \label{eq:convex_combination}
\end{equation}	
which represents the net effect of the fluctuating noise. Computing this stochastic average it can be shown that, to leading order in the noise strength $\gamma$, the state $\rho(t)$ follows a master equation of the following form (see Appendix~\ref{app:noise} for a sketch of this derivation) 
\begin{align}
    \frac{d}{dt}\rho(t)= -i[H(t),\rho(t)] -\gamma^2 \sum_{a}\int_0^tds \,D_{a}(t,s) \: \left[X_a,  [\widetilde{X}_{a}(s-t),\,\rho(t)]\,\right] \,. 
\end{align}
where $X_a:= \ket{a} \bra{\mathrm{f}} +\mathrm{h.c.}$ is a Hermitian operator that reflects the coupling of the noise to the system's Hamiltonian and $ \widetilde{X}_a(t)= \mathcal{U}^\dag(t)X_a\mathcal{U}(t)$, where $\mathcal{U}(t)$ is the time evolution under the unperturbed Hamiltonian $H(t)$. Notice that in the limit in which the noise correlations decay on a time-scale much faster than the one of the evolution under $H(t)$, we can assume $\widetilde{X}_a(s-t)\simeq\widetilde{X}_a(0)\simeq X_a$, which leads to
\begin{align}
    \frac{d}{dt}\rho(t)= -i[H(t),\rho(t)] +\sum_{a}2 \gamma^2 F_{a}(t) \left(X_a\rho(t)X_a-\frac{1}{2}\{X_a^2,\rho(t)\}\right)\,, \label{eq:master_equation}
\end{align}
with $F_{a}(t)=\int_0^tds \,D_{a}(t,s)$. 
This strongly resembles the Lindbladian noise terms arising from environment-induced decoherence, representing an instance of anisotropic Pauli noise.

\begin{figure}
    \centering
\includegraphics[width=0.9\linewidth]{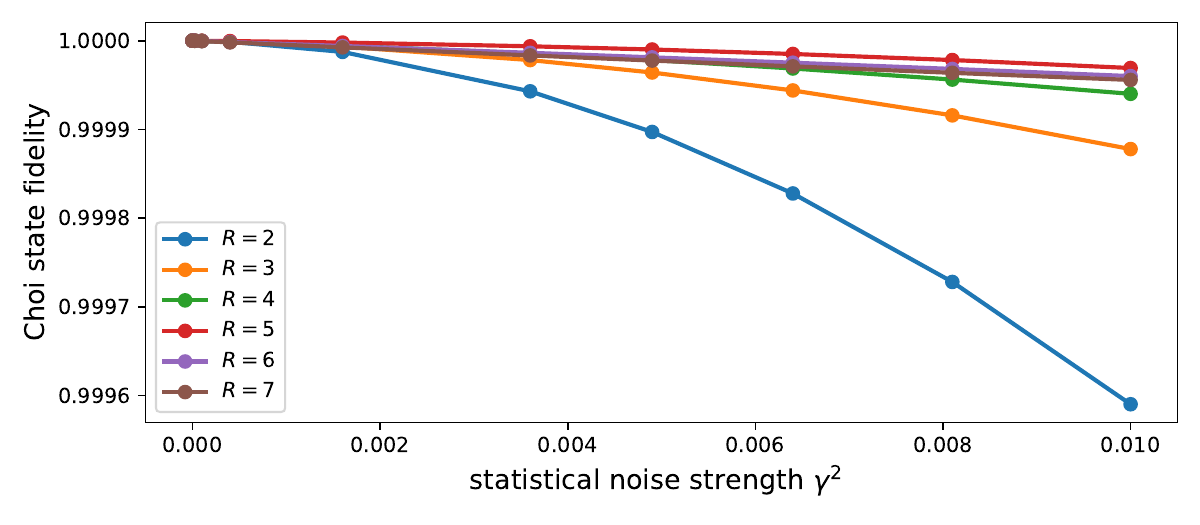}
    \caption{Simulation results for Lindbladian evolution according to Eq.~\eqref{eq:master_equation}. For simplicity, we consider the single-atom Hamiltonian~\eqref{eq:1qubit_hamiltonian} implementing in the adiabatic limit the qubit gate $\ketbra{0}{0} + e^{-i\pi/4}\ketbra{1}{1}$. The concrete evolution times used are $t_1=50, t_2=20$ as in Eq.~\eqref{eq:simple-loop-def}. We compare the noisy implemented map with the noiseless case ($\gamma =0$) on their respective restrictions to the computation subspace and treat these as quantum channels (though this is exact only in the adiabatic limit). As a distance measure we use the fidelity between the two normalized Choi states. For moderate loop radii $R < 6$, the fidelity increases with larger $R$, indicating improved noise resilience. The deviation for $R \geq 6$ might come from the fact that the assumption of timescale separation between the noise and the coherent evolution leading to Eq.~\eqref{eq:master_equation} breaks down for larger $R$, since for fixed evolution times, the rate of change of the coherent evolution becomes faster.}
    \label{fig:lindbladian-sim}
\end{figure}

In conclusion, any noise of the form as in Eq.~\eqref{eq:master_equation} can be interpreted as a convex combination of Gaussian control errors. As mentioned, we can expect each term in the convex combination~\eqref{eq:convex_combination} to be robust to the respective coherent noise realisation.
Therefore, also the mixture $\rho(t)$ should be robust to the noise model, now expressed in the form of the master equation~\eqref{eq:master_equation}.
In light of this, one may expect our holonomic scheme to be robust not only to errors in classical control, but also to families of environment-induced quantum noise. To confirm this reasoning, we have numerically simulated the master equation~\eqref{eq:master_equation} for the one-atom Hamiltonian with $F_a(t) \equiv \delta_{1,a}$ for different noise strengths $\gamma$ and different radii $R$ of the loop from Eq.~\eqref{eq:simple-loop-def}. As depicted in Fig.~\ref{fig:lindbladian-sim}, we observe that the Choi state fidelity between the noisy ($\gamma > 0$) and noiseless ($\gamma = 0$) implementations increases with larger loop radius $R$ for moderate radii ($R<6$), reflecting a behaviour analogous to the one encountered for coherent noise in Figure~\ref{fig:numerics_coh_error}.

\section{Conclusions and outlook} \label{sec:discussion}
In this work, we introduce a fully holonomic, scalable architecture for quantum computing well-suited for Rydberg atoms experiments. Specifically, we present a novel proposal to realize holonomic two-qudit entangling gates adiabatically in such a system for the first time. While our adiabatic approach requires longer implementation times compared to the gates described in, for example, Ref.~\cite{Levine19}, the fully holonomic nature of our gates suggests potential advantages in terms of robustness to certain types of errors, as we have carefully discussed.

Our analysis clarifies the nature of error resistance in holonomic gates, contributing to a deeper understanding of their robustness. 
In the differential geometric picture that we develop, we
demonstrate that these errors can be analyzed through the curvature of the vector bundle associated with the null subspaces of the Hamiltonian under varying parameters. By appropriately selecting loops in the parameter space, one can ensure that in most of the regions traversed by the loop the curvature is effectively zero. In this regard, our framework can be seen as being
analogous to paradigms of fault tolerance in quantum
computation that are captured in a differential geometric picure~\cite{gottesman2017fibre}, where noise resistance is achieved through the existence of a vector bundle with exactly zero curvature. 
We believe that this improved understanding of the nature of holonomic error robustness will be very valuable also for the study of other holonomic proposals. This link also further suggests that the expected robustness may go beyond errors in classical control.

The range of potential applications of the quantum gate set that we construct is broad. Compared to Ref.~\cite{Levine19}, our proposal enables the direct implementation of a substantially richer class of single-qubit and two-qubit gates, and it immediately extends to higher-dimensional qudits (as detailed in our equations above). This versatility makes our proposal particularly relevant for applications such as simulating \emph{non-Abelian lattice gauge theories}~\cite{LatticeGauge,GZCKZ22}. 

Another potentially significant application of our scheme is to implement physical gates within \emph{error corrected} architectures for quantum computing. Rydberg atom platforms have recently become prominent candidates for this type of protocols~\cite{bluvstein_quantum_2024}, mainly due to the long-range connectivity that they can achieve (both using reconfigurable tweezer arrays~\cite{xu_constant-overhead_2024} and exploiting the long distance Rydberg blockade interaction~\cite{pecorari_high-rate_2025}). This feature is particularly important for realizing \emph{high-rate quantum low-density parity-check} (qLDPC) codes \cite{GottesmanLDPC,PRXQuantum.2.040101} that have a number of very appealing parameters and features, except that they are necessarily non-local. In this context, the robustness of our gate implementation may be a precious tool to reach error-correction threshold values, despite their relatively slower implementation times.

More generally, the slow operation times of adiabaitic holonomic gates compared to resonant, pulse-based gate implementations need not preclude their usefulness in platforms with sufficiently long coherence times, such as neutral-atom systems with long-lived internal states. In this broader perspective, holonomic gates may represent a viable complementary approach for implementing robust quantum operations, particularly against certain classes of coherent control errors.
Beyond immediate performance considerations, holonomic quantum gates also exhibit a rich geometric structure, which continues to motivate their study from a theoretical standpoint and may inform future approaches to error-robust gate design.

To conclude, we would like to pose several questions for future research. 
First of all, we would like to stress that the parameter loop introduced in Eq.~\eqref{eq:simple-loop-def} and studied in our numerical results is by no means optimal in terms of its performance in an adiabatic algorithm. Most likely, the run-times of the protocol can be significantly improved by finding smoother loop shapes and by better adapting the evolution schedule to the gap. 
Further techniques from \emph{optimal control theory} could also be leveraged, 
including concepts of \emph{shortcuts to adiabaticity}. 
These have been already studied in this context in the theory of non-adiabatic holonomic computation: it would certainly be of impact to extend our insights also to this slightly different regime \cite{RevModPhys.91.045001,PhysRevApplied.15.024038}. 
%\textcolor{red}{[Cite some stuff]}

Another, possibly more fundamental, question that naturally arises is whether it is possible to design an adiabatic, holonomic gate within our framework that is implemented by a loop fully constrained to regions of negligible curvature: for instance, a loop that avoids 
the origin of our parameter space, where the curvature is maximum. Would this inevitably necessitate modifications to the effective Hamiltonian outlined in Eq.~\eqref{eq:1qubit_hamiltonian}? Such a gate could 
potentially offer enhanced robustness to fluctuations in control parameters, akin to the rigorous definition of fault tolerance described in Ref.~\cite{gottesman2017fibre}. It would also be exciting to make this connection more precise on the 
conceptual and rigorous levels.

Finally, would it be possible to identify more generic families of loops that naturally implement richer gate sets? In our present scheme any required (entangling) gate can be implemented by a concatenation of loops producing gates of the forms described in Eqs.~\eqref{eqs:our_gate_1and2qubits}. Would it be possible to find a single more elaborate loop that directly implements the required gate? Answering this question would require handling the time ordered exponential~\eqref{eq:time-ordered-int} in more complicated regimes than the ones we consider, for example with numerical optimization methods. It is the hope that the present work inspires such endeavors.

\section*{Acknowledgements}
We are grateful with Benjamin Schiffer and Ansgar Burchards for enlightening discussions. This work has been supported by the
German Federal Ministry for Education and Research (BMBF) under the projects 
FermiQP, DAQC, and MUNIQC-Atoms, the Munich Quantum Valley (K-8),
Berlin Quantum, the Quantum Flagship (PasQuans2, Millenion), 
the DFG (CRC 183), and the European Research Council
(DebuQC).

\bibliographystyle{unsrtnat}
\bibliography{refs}

\appendix
\section{Physical derivation of relevant Hamiltonian from dipole approximation}\label{sec:app_Hphysical}

The relevant single-atom Hamiltonian~\eqref{eq:1qubit_hamiltonian} is modelling the interaction of an atom with multiple monochromatic lasers. The $d+1$ lasers resonantly drive the transitions between $d+1$ atomic states $\{\ket{a}\}_{a =0}^d$  and an excited state $\ket{\mathrm f}$. The resonant interaction justifies treating the atom as a $(d+2)$-level atom with Hamiltonian
\begin{equation}
    H_{\text{atom}} = \sum_{a = 0}^d E_{a}\ketbra{a}{a} + E_{\mathrm f} \ketbra{\mathrm f}{\mathrm f}. 
\end{equation}
Each laser is described by a classical electric field 
\begin{equation}
   \vec{\mathcal E}_a =  \vec{\mathcal E}^{(0)}_a e^{-i\omega_a t} + h.c.
\end{equation}
where the resonant driving implies that $\omega_a = E_{\mathrm f} - E_a$. 
Under the dipole approximation in the general multipole expansion of the atom-field interaction, the interaction Hamiltonian takes the form $H_{\text{int}} = - \sum_{a=0}^d \vec{\mathrm d}\cdot\vec{\mathcal E}_a$, where $\vec{\mathrm d}$ is the dipole moment operator of the atom. The dipole operator only couples energy eigenstates of opposite spatial parity \cite{griffithsIntroductionQuantumMechanics2018}. Therefore, all its diagonal elements in the energy eigenbasis vanish $\bra{a}\vec{\mathrm d}\ket{a} = \bra{\mathrm f}\vec{\mathrm d}\ket{\mathrm f} =0$.  Assuming that $\ket{\mathrm f}$ has opposite parity from the rest of the energy states $\{\ket{a}\}_{a =0}^d$, the interaction Hamiltonian can be written as 
\begin{align}
    H_{\text{int}} &=  \sum_{a=0}^d  \sum_{a'=0}^d - \Bigl(\bra{\mathrm f}\vec{\mathrm d}\cdot\vec{\mathcal E}^{(0)}_a \ket{a'} e^{-i\omega_a t} +\bra{\mathrm f}\vec{\mathrm d}\cdot\vec{\mathcal E}^{(0)\ast}_a \ket{a'} e^{i\omega_a t}\Bigr)\ketbra{\mathrm f}{a'} + h.c.
\end{align}
The total Hamiltonian of the system is $H_{\text{atom}} + H_{\text{int}}$. At this point we consider a time-dependent change of basis defined by the unitary $U = e^{itH_{\text{atom}}}$. The Hamiltonian in this rotating frame then reads 
\begin{align}
    H &= U(H_{\text{atom}} + H_{\text{int}})U^\dagger +i\frac{\partial U}{\partial t} U^\dagger \\
    &= H_{\text{atom}} + U H_{\text{int}} U^\dagger +i \bigl(i H_{\text{atom}}\bigr) \\
    &= \sum_{a=0}^d  \sum_{a'=0}^d - \Bigl(\bra{\mathrm f}\vec{\mathrm d}\cdot\vec{\mathcal E}^{(0)}_a \ket{a'} e^{-i(\omega_a - \omega_{a'}) t} +\bra{\mathrm f}\vec{\mathrm d}\cdot\vec{\mathcal E}^{(0)\ast}_a \ket{a'} e^{i(\omega_a + \omega_{a'}) t}\Bigr)\ketbra{\mathrm f}{a'} + h.c. \quad  .
\end{align}
Next, we invoke the \textit{rotating wave approximation} where we coarse-grain the dynamics at the very fast optical frequencies. This means that we replace terms oscillating with a large frequency (as is $\omega = \omega_{a}+\omega_{a'}$) with their average value which is zero. So under the assumption that also $|\omega_a - \omega_{a'}|$  is sufficiently large for any $a \neq a'$ we get
\begin{align}
    H^{(\text{RWA})} &=  \sum_{a=0}^d - \bra{\mathrm f}\vec{\mathrm d}\cdot\vec{\mathcal E}^{(0)}_a \ket{a} \ketbra{\mathrm f}{a} + h.c. 
\end{align}
Defining the Rabi frequencies $\Omega^\ast_a = - \bra{\mathrm f}\vec{\mathrm d}\cdot\vec{\mathcal E}^{(0)}_a \ket{a} $ we recover the Hamiltonian ~\eqref{eq:1qubit_hamiltonian} investigated in this work. The interaction term $W\ketbra{d,d}{d,d}$ in the two-body Hamiltonian~\eqref{eq:2-atom-hamiltonian} models the Rydberg blockade interaction of two atoms trapped in optical potentials \cite{jakschFastQuantumGates2000}. The effect comes from the strong dipol-dipol interaction of two atoms excited to states of high principle quantum number $n$, called Rydberg states. This means that the state $\ket{d}$ in~\eqref{eq:2-atom-hamiltonian} corresponds to such a Rydberg level of the atom. In principle, the dipol-dipol interaction depends on the distance of the two atoms. Under the condition that the operation time of the gate is much smaller than the time scale of the external motion of the atoms, the interaction strength can be approximated with a constant value $W(\bm r) \equiv W$.

\section{Eigenvectors and eigenvalues of the relevant Hamiltonian}
\label{sec:app_eig_hamiltonian}
In this appendix we briefly summarize the eigenvectors of the relevant single-atom~\eqref{eq:1qubit_hamiltonian} and two-atom~\eqref{eq:2-atom-hamiltonian} Hamiltonians from the main text and additionally discuss the structure of the eigenvalues of the two-atom Hamiltonian. Furthermore, we investigate analytically and numerically the dependence on the Rydberg interaction strength $W$ of the smallest absolute energy value (i.e., the relevant energy gap $\Delta$ between the computational subspace and the nearest energy subspace). As already stated in the main text the eigenvalues of the single-atom Hamiltonian
\begin{equation*}
    H_0 = \sum_{a=0}^d\Omega_j \ketbra{a}{\mathrm{f}} + h.c.
\end{equation*}
are readily found to be $\pm\Om$, where $\Om^2=\sum_{a=0}^d|\Om_a|^2$, and $0$ ($d$-fold degenerate) with eigenvectors
\begin{equation*}
 \ket{\pm}=\Omega \ket{\mathrm{f}} \pm \sum_{a=0}^d \Omega_a \ket{a},
\end{equation*}
and
\begin{equation*}
    \ket{e_a}=\overline{\Omega}_a\ket d-\overline{\Omega}_d\ket a\,,
\end{equation*}
for all $0\le a\le d-1$.
In the case of the two-atom Hamiltonian 
\begin{equation*} 
H=1\otimes H_0+H_0\otimes1+ W \ket{d,d}\bra{d,d},
\end{equation*}
the subspace with energy eigenvalue $0$ is  $(d^2+1)$-dimensional and spanned by the following linearly independent sets of vectors 
\begin{equation*}
    \ket{w^0}=\ket{+,-}-\ket{-,+},
\end{equation*}
\begin{equation*}
    \ket{w^-_{ab}}=\ket{e_a,e_b}-\ket{e_b,e_a},
\end{equation*}
for all $0\le a<b\le d-1$, as well as the vectors
\begin{equation*}
\ket{w^+_{ab}}=\ket{e_a,e_b}+\ket{e_b,e_a}+\frac{\overline{\Omega}_a \overline{\Omega}_b}{\Omega_d^2}(\ket{+,-}+\ket{-,+})
\end{equation*}
for all $0\le a\le b\le d-1$. Vectors from different sets above are by construction mutually orthogonal. 

The remaining $4d+3$ non-zero eigenvalues were numerically found to be of two types. There are $2(2d-1)$ eigenvalues equal to the ones of the single-atom Hamiltonian $\pm \Omega$, with half of them being negative and the others positive. The rest correspond to the five real solutions of the equation
\begin{equation}
    x^5 -W x^4  -5\Omega^2 x^3+ W  (5\Omega^2 -2|\Omega_d|^2)x^2 +4\Omega^4 x - W(2|\Omega_d|^4 +4(\Omega^2-|\Omega_d|^2)^2) = 0.
    \label{eq:deg5_polynomial}
\end{equation}
Since the coefficients are composed of $|\Omega_a|$,  with $0\leq a \leq d$, and $W$, any non-zero eigenvalue depends only on the absolute values $|\Omega_a|$ and consequently the relevant energy 
gap does as well. When $W$ approaches zero, Eq.~\eqref{eq:deg5_polynomial} reduces to $x(x^2-\Omega^2)(x^2-4\Omega^2) =0$. In this limiting case $x_0 = 0$ is a solution, while in the general case Eq.~\eqref{eq:deg5_polynomial} has non-vanishing solutions. Since the zeros of a polynomial depend continuously on its coefficients, we conclude that the gap $\Delta$ interpolates smoothly between zero and a finite value as $W$ is varied.

\begin{figure}
    \centering
        \begin{subfigure}[b]{0.48\textwidth}
        \centering
    \includegraphics[width=0.96\textwidth]{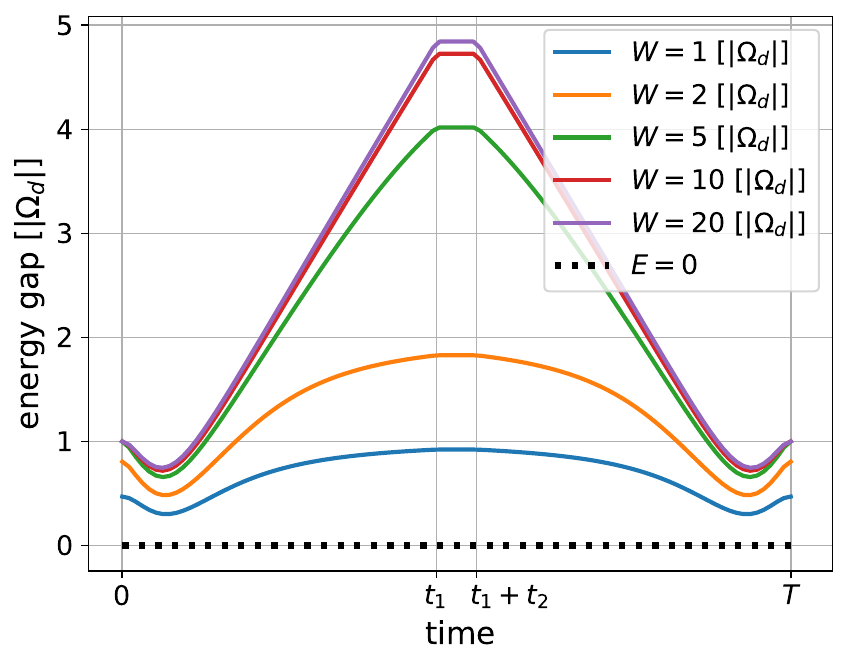}
        \caption{}
    \end{subfigure}
    \begin{subfigure}[b]{0.48\textwidth}
        \centering
        \includegraphics[width=\textwidth]{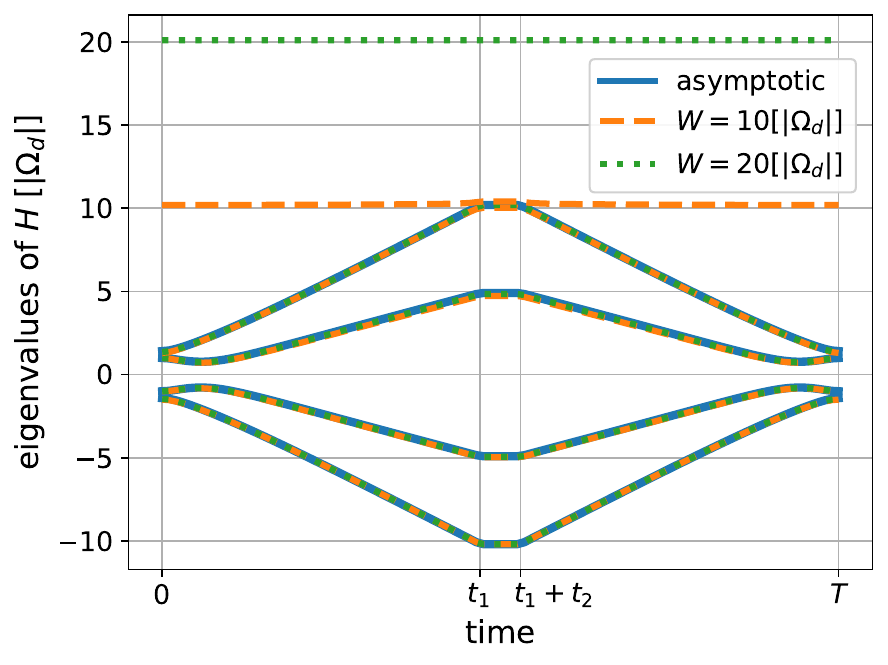}
        \caption{}
    \end{subfigure}
    \caption{a) Dependence of the gap $\Delta$ of the two-qubit Hamiltonian on the Rydberg interaction strength $W$ along the path from Fig.~\ref{fig:pacman_loop}. The value of the gap saturates already at $W \approx10[|\Omega_d|]$. Higher Rydberg interaction values will thus not lead to a smaller adiabatic error and an improved performance of the gate. b) The five eigenvalues of the two-atom Hamiltonian corresponding to the solutions of Eq.~\eqref{eq:deg5_polynomial} are plotted for two large Rydberg interaction values. They are already very well described by the asymptotic solutions anticipated in the main text: The eigenvalue $\approx W$ is apparent, as well as the two positive solutions and the corresponding symmetric negative ones as described by Eq.~\eqref{eq:sol_asym_eigenvalues}.}
    \label{fig:gaps_Rydinter_strength}
\end{figure}

To say more, we numerically investigated the dependence of the energy gap $\Delta$ with respect to the Rydberg interaction $W$ along the loop parametrized as in Fig.~\ref{fig:pacman_loop}. We observe in Fig.~\ref{fig:gaps_Rydinter_strength}a that the gap value saturates with increasing Rydberg interaction strength already at $W\approx10\,[|\Omega_d|]$. This tells us on a heuristic level that it is not necessary to increase the interaction beyond this value because it will not result in a smaller adiabatic error. 
To obtain an analytic approximation to this value of the gap we consider the case of very strong Rydberg interaction, i.e., $W \gg 1$ and $W \gg \Omega$. One of the eigenvectors will in this limit approach $|d,d\rangle$ with eigenvalue $W$. The remaining four are obtained considering Eq.~\eqref{eq:deg5_polynomial} to leading order in $1/W$, to get
\begin{equation}\label{eq:deg5_polynomial-approx}
       x^4  -  (5\Omega^2 -2|\Omega_d|^2)x^2  + (2|\Omega_d|^4 +4(\Omega^2-|\Omega_d|^2)^2) = 0. 
\end{equation}
For convenience we define $D^2 := \sum_{a=0}^{d-1}|\Omega_a|^2 = \Omega^2 -|\Omega_d|^2$. Then the solutions of Eq.~\eqref{eq:deg5_polynomial-approx} are
\begin{equation}
\begin{aligned}
       %& \pm \frac{1}{\sqrt{2}}\Bigl(3|\Omega_d|^2 + 5D^2 \pm \sqrt{(3|\Omega_d|^2 + 5D^2)^2 - 4(2|\Omega_d|^4 +4D^4))}\Bigr)^{1/2}\\
     % &= \pm \frac{1}{\sqrt{2}}\Bigl(5\Omega^2 -2|\Omega_d|^2 \pm \sqrt{25\Omega^4 -20\Omega^2|\Omega_d|^2 +4|\Omega_d|^4 - 8|\Omega_d|^4 -16\Omega^4 +32|\Omega_d|^2\Omega^2 - 16|\Omega_d|^4)}\Bigr)^{1/2} \\
      & \pm \frac{1}{\sqrt{2}}\Bigl(3|\Omega_d|^2 + 5D^2\pm \sqrt{|\Omega_d|^4 +30D^2|\Omega_d|^2 +9 D^4}\Bigr)^{1/2}.
\end{aligned}
\label{eq:sol_asym_eigenvalues}
\end{equation}
By continuity,  we again expect these asymptotic eigenvalues to describe reasonably well the solutions of Eq.~\eqref{eq:deg5_polynomial} for large but finite $W$.\@ Indeed, we numerically find (see Fig.~\ref{fig:gaps_Rydinter_strength}b)  that all the eigenvalues are very well captured by the behaviour found above for the asymptotic eigenvalues already for $W = 10,20 \,[|\Omega_d|]$. 
The smallest solution in Eq.~\eqref{eq:sol_asym_eigenvalues} in absolute value is the relevant gap. Hence, for a Rydberg interaction of around $W \geq 10\,[|\Omega_d|]$, it is well described by 
\begin{equation*}
    \Delta \approx  \frac{1}{\sqrt{2}}\Bigl(3|\Omega_d|^2 + 5D^2 - \sqrt{|\Omega_d|^4 +30D^2|\Omega_d|^2 +9 D^4}\Bigr)^{1/2}.
\end{equation*}
This is lower-bounded by
\begin{equation*}
    \Delta \geq\sqrt{\frac{3|\Omega_d|^2}2\left(1 - \sqrt{1-\frac{16}{43}}\right)} \geq \frac{1}{2}|\Omega_d|\,,
\end{equation*}
which allows us to conclude that, for sufficiently large positive $W$, the gap stays open for any value of the Hamiltonian parameters $\Omega_a$.

\section{Detailed expressions for the two-qudit gate} \label{app:formulae}
In Sec.~\ref{sec:methods-2qudits} we have stated that the Gram matrix $g$ and the connection tensor $A$ of the two-qudit system are block diagonal with respect to the subspaces~\eqref{eq:null-ev-2qudit_1}, \eqref{eq:null-ev-2qudit_2} and~\eqref{eq:null-ev-2qudit_3}. This stems from the following observations. First notice that $\ket{w_{ab}^+(\bla)}$ and $\partial_c\ket{w_{ab}^+(\bla)}$ are symmetric with respect to exchanging the two qudits, while $\ket{w^0}$ and $\ket{w_{ab}^-(\bla)}$ are anti-symmetric. It follows that the former are orthogonal to the latter, and that $g$ and $A$ have a separate block corresponding to the subspace spanned by $\ket{w_{ab}^+(\bla)}$. To show the existence of the further two blocks, we then just need to prove $\braket{w^0|w_{ab}^-(\bla)}=\braket{w^0|\partial_c w_{ab}^-(\bla)}=0$. This follows directly from the fact that $\braket{\pm|e_a(\bla)}=0$, as one can find from the single-qudit analysis.

In the rest of this section we will list the analytic expressions for the blocks of the Gram matrix $g$ and of the connection tensor $A$ of the two-qudit system, which we have deferred from the main text for better readability. In particular, we have that the quantity $g^+$ discussed in Sec.~\ref{sec:methods-2qudits} has the explict form
\begin{multline*}
    g^{+}_{(a,b),(k,l)}(\bla):=\braket{w_{ab}^+(\bla)|w^+_{kl}(\bla)}\\
    =2 |\Omega_d|^4 \left[ \delta_{ak} \delta_{bl}   +\frac{1}{|\Omega_d|^2}(\Omega_a \overline{\Omega}_k \delta_{bl} +\delta_{ak} \Omega_b \overline{\Omega}_l  )  +\frac{\alpha}{|\Omega_d|^4}  \Omega_a \Omega_b \overline{\Omega}_k\overline{\Omega}_l\right] + (k \leftrightarrow l)\,,
\end{multline*}
where $\alpha=1+2\frac{\Omega^4}{|\Omega_d|^4}$, for all $0\le a\le b\le d-1$ and $0\le k\le l\le d-1$. It can be determined that its inverse has a similar form, namely
\begin{equation}
    g^{+\,(a,b),(k,l)}(\bla)=\frac{1}{8|\Omega_d|^{4}}  \left[ \delta_{ak} \delta_{bl}  -\frac{1}{\Omega^2}(\Omega_a \overline{\Omega}_k \delta_{bl} +\Omega_b \overline{\Omega}_k \delta_{al} ) + \frac{\beta}{\Omega^4}\Omega_a \Omega_b \overline{\Omega}_k\overline{\Omega}_l\right] + (k \leftrightarrow l)\,, \nonumber
\end{equation}
where $\beta= ( |\Omega_d|^4  -4\Omega^2 |\Omega_d|^2 +2\Omega^4)(3  |\Omega_d|^4 -4 \Omega^2 |\Omega_d|^2 +2\Omega^4)^{-1}$, for all $0\le a\le b\le d-1$ and $0\le k\le l\le d-1$. For what concerns the connection, we have that for $0\le c\le d-1$
\begin{multline*}
        A^{+}_{(k,l),(a,b),c}(\bla):=\braket{w^+_{kl}(\boldsymbol{\lambda})|\partial_c w^+_{ab}(\boldsymbol{\lambda})}\\=2 |\Omega_d|^2(\delta_{cb} \delta_{ak} \Omega_l + \delta_{ca} \delta_{bl} \Omega_k +\delta_{ca} \delta_{bk} \Omega_l + \delta_{cb} \delta_{al} \Omega_k )\\
        + 4 \,  \Omega_k \Omega_l(\delta_{ca} \overline{\Omega}_b + \delta_{cb} \overline{\Omega}_a) \,  \left(1+2\frac{\Omega^4}{|\Omega_d|^4}\right)\\
        + 4 \, \frac{\Omega_k \Omega_l \overline{\Omega}_a \overline{\Omega}_b} {|\Omega_d|^4} \: (3\overline{\Omega}_c +\Omega_c)\, \Omega^2\,,
\end{multline*}
while for 
 $d\le c\le2d-1$
\begin{multline*}
        A^{+}_{(k,l),(a,b),c}(\bla):=\braket{w^+_{kl}(\boldsymbol{\lambda})|\partial_c w^+_{ab}(\boldsymbol{\lambda})}\\=-\iu\,2 |\Omega_d|^2(\delta_{c-d,b} \delta_{ak} \Omega_l + \delta_{c-d,a} \delta_{bl} \Omega_k +\delta_{c-d,a} \delta_{bk} \Omega_l + \delta_{c-d,b} \delta_{al} \Omega_k )\\
        -\iu\,4 \,  \Omega_k \Omega_l(\delta_{c-d,a} \overline{\Omega}_b + \delta_{c-d,b} \overline{\Omega}_a) \,  \left(1+2\frac{\Omega^4}{|\Omega_d|^4}\right)\\
        + 4 \, \frac{\Omega_k \Omega_l \overline{\Omega}_a \overline{\Omega}_b} {|\Omega_d|^4} \: (3\iu\overline{\Omega}_{c-d} -\iu\Omega_{c-d})\, \Omega^2\,.
\end{multline*}
Here, the first expression corresponds to taking derivatives with respect to the real part of $\Omega_c$, while the second refers to taking derivatives with respect to the imaginary part, consistently with the definition $\boldsymbol{\lambda}=(\mathrm{Re}\, \Omega_0,\dots, \mathrm{Re}\, \Omega_{d-1},$ $\mathrm{Im} \,\Omega_0,\dots, \mathrm{Im} \,\Omega_{d-1})$.

We are now ready to combine these results to compute the connection with a raised index. After some algebra we find
\begin{multline*}
    \dot{\lambda}^\mu A^{+ \, (k,l)}{}_{(a,b),\mu}(\bla)=  
    \frac{1}{2\Omega^2} (\delta_{ka} \Omega_l \dot{\overline{\Omega}}_b+ \delta_{la} \Omega_k \dot{\overline{\Omega}}_b+ \delta_{kb} \Omega_l \dot{\overline{\Omega}}_a + \delta_{lb} \Omega_k \dot{\overline{\Omega}}_a)\\
    +\Omega_k \Omega_l \overline{\Omega}_a \overline{\Omega}_b \: \frac{\sum_c(3 \dot{\Omega}_c \overline{\Omega}_c +\dot{\overline{\Omega}}_c\Omega_c)}{\Omega^2(3|\Omega_d|^4-4\Omega^2|\Omega_d|^2+2\Omega^4)}\\
    +\Omega_k \Omega_l (\dot{\overline{\Omega}}_a \overline{\Omega}_b + \overline{\Omega}_a \dot{\overline{\Omega}}_b)\, \frac{2|\Omega_d|^2}{\Omega^2(3|\Omega_d|^4-4\Omega^2|\Omega_d|^2+2\Omega^4)}\,.
\end{multline*}

\section{Insights on fluctuating noise}\label{app:noise}

In this section we want to describe the impact of adding the stochastic noise term~\eqref{eq:omega+xi} to the system's Hamiltonian~\eqref{eq:1qubit_hamiltonian}. To do this we will follow the derivation of Ref.~\cite{adler_collapse_2007}.
Consider a time evolution governed by the stochastic Schrödinger equation
\begin{equation}
    \frac{d}{dt} \ket{\psi(t)}=-i\left(H(t) + \gamma \sum_a \xi_a(t) X_a\right)\ket{\psi(t)} \label{eq:SSE}
\end{equation}
where $H(t)$ is any time-dependent Hamiltonian, perturbed by a fluctuating noise term depending on the Hermitian operators $X_a$, $a=1,\dots ,d$. Here, $\gamma$ is a real parameter determining the noise strength and $\xi_a(t)$ are Gaussian stochastic processes which satisfy
\begin{equation}
    \mathbb{E}[\xi_a(t)]=0\,, \hspace{10mm} \mathbb{E}[\xi_a(t)\xi_b(s)]=D_{ab}(t,s). \label{eq:correlation-function}
\end{equation}
As they are Gaussian processes, the covariance function $D_{ab}(t,s)$ fully determines the processes $\xi_a(t)$, in particular determining the time scale on which they display correlations. We can assume $\xi_a(t)$ to be almost always continuous. For simplicity, we will later assume that the noise processes are uncorrelated for different $a=1,\dots ,d$, that is $D_{ab}(t,s)=\delta_{ab}D_a(t,s)$. But for now we will make derivations in this slightly more general setting. 

A formal solution of Eq.~\eqref{eq:SSE} can be 
derived in the interaction picture with respect to $H(t)$. Consider indeed the unitary time evolution operator given by the following time ordered exponential
\begin{equation*}
    \mathcal{U}(t)=\mathcal{T}\exp\left(-i \int_0^t ds\, H(s)\right)\,.
\end{equation*}
If we define the interaction picture state vector as $\ket{\widetilde{\psi}(t)}=\mathcal{U}^\dag(t)\ket{\psi(t)}$, then this state vector  clearly satisfies the equation of motion
\begin{equation*}
    \frac{d}{dt} \ket{\widetilde{\psi}(t)}=-i \gamma \sum_a \xi_a(t) \widetilde{X}_a(t)\ket{\widetilde{\psi}(t)}\,,
\end{equation*}
with $ \widetilde{X}_a(t)= \mathcal{U}^\dag(t)X_a\mathcal{U}(t)$, which admits the formal solution
\begin{equation*}
    \ket{\widetilde{\psi}(t)}=\mathcal{T}\exp\left(-i\gamma\int_0^t ds\, \sum_a \xi_a(s) \widetilde{X}_a(s)\right) \ket{\psi(0)}\,.
\end{equation*}
Substituting this into $\ket{\psi(t)}=\mathcal{U}(t)\ket{\widetilde{\psi}(t)}$ and expanding the exponential to first order in $\gamma$ we have
\begin{equation}
    \ket{\psi(t)}=\mathcal{U}(t)\left(\id -i\gamma\int_0^t ds\, \sum_a \xi_a(s) \widetilde{X}_a(s) +\mathcal{O}(\gamma^2)\right) \ket{\psi(0)}\,.\label{eq:sol_psi_t}
\end{equation}
From the point of view of the quantum state's density matrix, we equivalently have
\begin{align}
    \ket{\psi(t)}\!\!\bra{\psi(t)}&=\mathcal{U}(t) \ket{\psi(0)}\!\!\bra{\psi(0)}\mathcal{U}^\dag(t) \nonumber\\
    \nonumber&\hspace{15mm}-i\gamma \int_0^t ds\, \sum_a \xi_a(s)\,\mathcal{U}(t) \,[\widetilde{X}_a(s),\ket{\psi(0)}\!\!\bra{\psi(0)}]\,\mathcal{U}^\dag(t)+\mathcal{O}(\gamma^2) \\[2mm]
    &=\mathcal{U}(t) \ket{\psi(0)}\!\!\bra{\psi(0)}\mathcal{U}^\dag(t) \nonumber\\
    \nonumber&\hspace{15mm}-i\gamma \int_0^t ds\, \sum_a \xi_a(s)\,[\widetilde{X}_a(s-t),\,\mathcal{U}(t) \ket{\psi(0)}\!\!\bra{\psi(0)}\mathcal{U}^\dag(t)\,]+\mathcal{O}(\gamma^2) \\[2mm]
    &=\mathcal{U}(t) \ket{\psi(0)}\!\!\bra{\psi(0)}\mathcal{U}^\dag(t) \nonumber\\
    &\hspace{15mm}-i\gamma \int_0^t ds\, \sum_a \xi_a(s)[\widetilde{X}_a(s-t),\,\ket{\psi(t)}\!\!\bra{\psi(t)}\,]+\mathcal{O}(\gamma^2) \,,\label{eq:sol_psi_psi_t}
\end{align}
where in the last step we have used the fact that, according to Eq.~\eqref{eq:sol_psi_t}, $\ket{\psi(t)}=\mathcal{U}(t) \ket{\psi(0)}+\mathcal{O}(\gamma)$. Eq.~\eqref{eq:sol_psi_psi_t} in particular implies the following fact (that we will use later) about the functional derivative of the state with respect to the noise process
\begin{equation*}
    \frac{\delta}{\delta \xi_a(s)} \ket{\psi(t)}\!\!\bra{\psi(t)} = -i\gamma \, \theta(t-s) \, [\widetilde{X}_a(s-t),\,\ket{\psi(t)}\!\!\bra{\psi(t)}\,]+\mathcal{O}(\gamma^2) \,.
\end{equation*}
Here, $\theta$ is the Heaviside step function.
Indeed, we would now like to apply the Furutsu-Novikov formula to compute the correlation $\mathbb{E}[\xi_a(t)\ket{\psi(t)}\!\!\bra{\psi(t)}]$. The Furutsu-Novikov formula~\cite{furutsu_statistical_1964,novikov_functionals_1964} states that for any functional $F[\xi(t)]$ of a Gaussian process $\xi(t)$ we have
\begin{equation*}
\mathbb{E}\left[\xi_a(t)F[\xi(t)]\right]=\sum_b\int ds \, D_{ab}(t,s) \: \mathbb{E}\left(\frac{\delta}{\delta \xi_b(s)}F[\xi(t)]\right)\,,
\end{equation*}
where $D_{ab}(t,s)$ is defined as in Eq.~\eqref{eq:correlation-function}. If we now choose $F[\xi(t)]=\ket{\psi(t)}\!\!\bra{\psi(t)}$, we immediately have
\begin{align}
\mathbb{E}\left[\xi_a(t)\ket{\psi(t)}\!\!\bra{\psi(t)}\right]=-i\gamma \sum_b\int_0^tds \, D_{ab}(t,s) \: [\widetilde{X}_a(s-t),\,\mathbb{E}\ket{\psi(t)}\!\!\bra{\psi(t)}\,] + \mathcal{O}(\gamma^2)\,.\label{eq:correlation-xi-psipsi}
\end{align}
Let us now finally consider the state that is obtained by sampling many instances of the noise process $\xi$ and averaging over the corresponding state vector trajectories $t\mapsto \ket{\psi(t)}$. This state is represented by the density operator
\begin{equation*}
    \rho(t)=\mathbb{E}\ket{\psi(t)}\!\!\bra{\psi(t)}\,.
\end{equation*}
We would like to derive the master equation that describes the mixed state evolution of $\rho(t)$. To find this let us combine our previous results and observe that
\begin{align}
\nonumber
    \frac{d}{dt}\rho(t)&= \mathbb{E} \left( \frac{d}{dt}\ket{\psi(t)}\!\!\bra{\psi(t)} \right)\\
    \nonumber&=\mathbb{E} \left( -i[H(t),\ket{\psi(t)}\!\!\bra{\psi(t)} ] -i\gamma \sum_a [\xi_a(t)X_a,\ket{\psi(t)}\!\!\bra{\psi(t)} ]\right) \label{eq:insert-vonneumann}\\
    &= -i[H(t),\mathbb{E}\ket{\psi(t)}\!\!\bra{\psi(t)} ]-i\gamma \sum_a \left[X_a,\mathbb{E}[\xi_a(t)\ket{\psi(t)}\!\!\bra{\psi(t)}]\,  \right]\\
    \nonumber
    &= -i[H(t),\rho(t)] -\gamma^2 \sum_{ab}\int_0^tds \,D_{ab}(t,s) \: \left[X_a,  [\widetilde{X}_b(s-t),\,\rho(t)]\,\right] + \mathcal{O}(\gamma^3)\,,
\end{align}
where in step~\eqref{eq:insert-vonneumann} we have just inserted the Von Neumann equation corresponding to the Schr\"odinger equation~\eqref{eq:SSE} and in the last step we have used the result~\eqref{eq:correlation-xi-psipsi}.

To conclude, notice that if the noise fluctuations become rapidly uncorrelated, then $D_{ab}(t,s)$ decays quickly when $s$ is not close to $t$. If the time-scale of this noise correlation decay is much faster than the typical time-scales of the evolution $H(t)$, then we can approximate $s-t\simeq 0$ and find $\widetilde{X}_a(s-t)\simeq X_a$. If we also assume, as anticipated above, that the noise processes $\xi_a(t)$ are uncorrelated for different $a$, this leads to a master equation in Lindblad form 
\begin{align}
    \frac{d}{dt}\rho(t)&=-i[H(t),\rho(t)] -\gamma^2 \sum_{a}F_{a}(t) \: \left[X_a,  [X_a,\,\rho(t)]\,\right] + \mathcal{O}(\gamma^3)\nonumber\\
    &= -i[H(t),\rho(t)] +\sum_{a}2 \gamma^2 F_{a}(t) \left(X_a\rho(t)X_a-\frac{1}{2}\{X_a^2,\rho(t)\}\right)+ \mathcal{O}(\gamma^3)\nonumber\,,
\end{align}
with time-dependent coefficients $F_{a}(t)=\int_0^tds \,D_{a}(t,s)$.

\end{document}